\documentclass[11pt]{article}

\usepackage[margin=72pt]{geometry}
\usepackage{mathtools}
\usepackage{setspace} 
\usepackage{tikz}
\usetikzlibrary{calc}
\usepackage{cite}
\usepackage{algorithm} 
\usepackage{algpseudocode}
\usepackage{float}
\usepackage{caption}
\usepackage{subcaption}
\usepackage{amsmath,amssymb,amsfonts,amsthm,verbatim}
\usepackage{graphicx}
\usepackage[justification=centering]{caption}
\usepackage{textcomp}
\usepackage{xcolor}
\usepackage{pgfplots}
\pgfplotsset{compat=1.7,
emphasize/.code args={#1:#2with#3}{
    \pgfplotsextra{
            \draw[fill=#3] ({axis cs:#1,0} |- {axis description cs:0,0}) 
            rectangle ({axis cs:#2,0} |- {axis description cs:0,1});
        }
    }
}
\usepackage[utf8]{inputenc}
\usepackage[T1]{fontenc}

\def\BibTeX{{\rm B\kern-.05em{\sc i\kern-.025em b}\kern-.08em
    T\kern-.1667em\lower.7ex\hbox{E}\kern-.125emX}}

\DeclareMathOperator{\rank}{rk}
\DeclareMathOperator{\supp}{supp}

\DeclareMathOperator{\wt}{wt}
\DeclareMathOperator{\rows}{Rows}
\newcommand{\prob}{\mathbb{P}}
\DeclareMathOperator{\pir}{PIR}
\DeclarePairedDelimiter\ceil{\lceil}{\rceil}
\DeclarePairedDelimiter\floor{\lfloor}{\rfloor}
\newcommand\compequi{\stackrel{\mathclap{\normalfont\mbox{c}}}{\approx}}
\newcommand{\F}{\mathbb{F}}

\newtheorem{definition}{Definition}
\newtheorem{lemma}{Lemma}
\newtheorem{theorem}{Theorem}
\newtheorem{proposition}{Proposition}

\newtheorem{example}{Example}
\newtheorem{corollary}{Corollary}

\newtheorem{remark}{Remark}

\newif\ifcomment
\commenttrue
\commentfalse 

\begin{document}

\title{CB-cPIR: Code-Based Computational Private Information Retrieval}

\author{Camilla Hollanti$^1$ \and Neehar Verma$^1$}
\date{%
    $^1$ Department of Mathematics and Systems Analysis, Aalto University
}

\maketitle

\let\thefootnote\relax\footnotetext{This work was supported by the Finnish Research Council (grant \#351271) and by the MSCA Doctoral Networks 2021, HORIZON-MSCA-2021-DN-01 (ENCODE, grant \#101072316). This work was done in part while the second author was visiting the Simons Institute for the Theory of Computing at the University of  California, Berkeley.\\ 
A preliminary version of this article was published in the Proceedings of the 2024 IEEE International Symposium on Information Theory (ISIT) \cite{verma2024isit}.}

\begin{abstract}
A private information retrieval (PIR) scheme is a protocol that allows a user to retrieve a file from a database without revealing the identity of the desired file to a curious database. Given a distributed data storage system, efficient PIR can be achieved by making assumptions about the colluding capabilities of the storage servers holding the  database. If these assumptions turn out to be incorrect, privacy is lost. In this work, we focus on the worst-case assumption: full collusion or, equivalently, viewing the storage system virtually as a single honest-but-curious server. We present \emph{CB-cPIR}, a single-server code-based computational private information retrieval (cPIR) scheme that derives security from code-based cryptography. Specifically, the queries are protected by the hardness of decoding a random linear code.  The scheme is heavily inspired by the pioneering code-based cPIR scheme proposed by Holzbaur, Hollanti, and Wachter-Zeh in [Holzbaur \emph{et al.}, ``Computational Code-Based Single-Server Private Information Retrieval'', 2020 IEEE ISIT] and fixes the vulnerabilities of the original scheme arising from highly probable rank differences in submatrices of the user's query. Recently, a new vulnerability was observed in [Lage, Bartz, ``On the Security of a Code-Based PIR Scheme''], a simple modification to the scheme now fixes this vulnerability. For further validation of our scheme, we draw comparisons to the state-of-the-art lattice-based cPIR schemes.
\end{abstract}

\section{Introduction}

Private information retrieval (PIR) was first introduced by Chor \textit{et al.} in \cite{chor1995private, PIR} with the aim of enabling users to access data from a database or, more generally, from a distributed storage system while concealing the identity of the requested information from potentially untrusted servers. A trivial way to guarantee information-theoretically secure PIR is to download the entire database. Modern data storage systems may often contain a large number of big (\emph{e.g.}, multimedia) files and the trivial solution is infeasible in practice. More practical solutions that attempt to incur minimal communication overhead and related  capacity results for \emph{information-theoretically secure} PIR schemes are presented in  \cite{sun2017repcap,sun2017capacity, freij2017private, sun2018capacity, banawan2018capacity, tian2019capacity,holzbaur2022tit}. To enable information-theoretic privacy these works assume that the distributed storage system consists of sufficiently large subsets of non-colluding servers. In practice, it may be difficult to decide for an appropriate level of collusion protection, and the more one protects, the more penalty there is in terms of the achievable PIR rates. Moreover, if too many servers collude, user privacy might be lost. For this reason, considering a single server or, equivalently, full collusion becomes interesting.
In this case,  information-theoretic privacy can only be achieved by downloading all the files. As a more practical alternative, \emph{computationally secure} schemes have been examined in several works. Certain schemes, \emph{e.g.}, \cite{kushilevitz1997replication, lipmaa2005oblivious, gentry2005single},
make use of computationally hard problems in the realm of classical computers, such as the quadratic residuosity problem. Such schemes will be rendered insecure when quantum computing matures, since the underlying hard problems can be efficiently solved using quantum algorithms.

\subsection{Related work and contributions} 

In the realm of post-quantum security, both lattice-based \cite{Peikert} and code-based cryptography \cite{weger2024survey} have emerged as promising avenues.

\paragraph{Lattice-based PIR:} In \cite{aguilar2007lattice}, an efficient lattice-based computational PIR scheme was proposed. Although this approach initially appeared robust, a practical vulnerability was revealed in \cite{liu2016cryptanalysis}, specifically targeting databases with a limited number of elements. However, such a limitation may not pose a significant threat, given the prevalent use of databases with a large number of elements.

The introduction of the first fully homomorphic encryption (FHE) scheme in \cite{gentry2009fully} marked a breakthrough in post-quantum cryptography and cryptography in general. Subsequently, FHE was leveraged to construct a general PIR scheme in \cite{yi2012single}. Several other PIR schemes based on FHE are presented in \cite{kiayias2015optimal,lipmaa2017simpler,gentry2019compressible}. Schemes based on FHE offer computationally secure PIR, but may often come at the cost of a high computational complexity. In this work, we compare the proposed \emph{CB-cPIR} scheme to two state-of-the-art lattice-based PIR schemes --- \emph{XPIR} and \emph{SimplePIR} --- that address the challenge of high computational complexity.

The XPIR \cite{aguilar2016xpir} scheme is based on the ring learning with errors problem (RLWE) \cite{RLWE} that combats the problem of high computational costs by utilizing an encryption scheme, which is just an additively-homomorphic building block of the FHE scheme in \cite{FHE-RLWE}. Moreover, polynomial multiplications are optimized using typical number-theoretic tools.

SimplePIR \cite{SimplePIR} in contrast to XPIR is based on the standard learning with errors (LWE) problem. The simplicity of LWE-based encryption allows reductions in computational costs by avoiding the need for polynomial multiplications. Using the weaker assumption of plain LWE comes with the drawback of high communication cost, which is mitigated by server-side preprocessing and distribution of a hint, which is reusable over multiple queries.

\paragraph{Code-based PIR:} In code-based cryptography, the goal is to use a structured code, \emph{e.g.} the McEliece scheme \cite{McEliece} with a binary Goppa code, which is difficult to distinguish from a random linear code. The security of the scheme is based on the hardness of decoding a random linear code, which is known to be NP-hard \cite{Berlekamp1978}.

The construction proposed in \cite{holzbaur2020isit} introduced \textbf{the first code-based PIR scheme}, referred to as the \emph{HHW~scheme} throughout this paper. In the HHW scheme, the server is queried using a matrix comprising intentionally corrupted codewords selected from a random linear code. The confidentiality of the desired file index is maintained through specifically crafted errors embedded in the query matrix. Upon receiving the server's response, decoding exposes the errors, and projection onto a relevant vector subspace unveils the desired file. As the locations of these errors were initially picked by the user, they simply need to do erasure decoding, making the scheme \emph{feasible}. This also provides the luxury that there is \emph{no need for the query code to be a structured code}. Hence, the scheme can genuinely rely on a random code providing the afore-mentioned \emph{security} guarantees due to the hardness of the decoding problem.   

Notably, the HHW scheme, with carefully chosen parameters, achieves PIR rates comparable to the computational PIR schemes presented in \cite{aguilar2007lattice,yi2012single}. For the proposed parameters in \cite[Section III.4]{aguilar2007lattice} the computational complexity is seen to be the complexity of matrix multiplications over the field $\mathbb{F}_{2^{60} + 325}$. For the HHW scheme with parameters achieving similar retrieval rates, the computational complexity is approximately equal to the multiplication of matrices of similar size over a significantly smaller field $\mathbb{F}_{2^{29}}$. Another attractive feature of the HHW scheme lies in its ability to perform calculations over binary extension fields. Despite its merits, the security of the HHW scheme was questioned in \cite{bordage2020privacy}. The identified vulnerability enables an attacker to discern the secret by observing rank differences in submatrices of the query; we will refer to this attack as \emph{subquery attack}.

In \cite{Alfarano2023survey} the authors develop a code-based framework, which formalizes several single-server PIR schemes. In this framework it is seen that any PIR scheme similar to the HHW scheme is susceptible to the subquery attack. The authors in \cite{bodur2023ring} circumvent this attack by using non-free codes over rings. These non-free codes are constructed by applying the Chinese Remainder Theorem to codes that are so-called non-Hensel lifts  \cite[Section IV, Corollary 7]{bodur2023ring}. This ring-based PIR protocol can achieve retrieval rates no more than $1/2n$, where $n$ determines the security level of the protocol. The scheme presented in this paper has no such limitation and can therefore outperform the ring-based scheme in terms of communication costs. 

\paragraph{Main contributions:} The proposed CB-cPIR scheme resurrects the HHW scheme by providing a remedy against the subquery attack and consequently to any similarly constructed scheme that is susceptible to this form of an attack. Furthermore, CB-cPIR preserves all the merits of the HHW scheme while now also ensuring privacy. Preliminary results were presented at ISIT 2024 \cite{verma2024isit}. Here, the following extensions are provided:
\begin{itemize}
\item A more comprehensive background on both code-based cryptography and single-server PIR is given.
\item A new attack is identified in Sec. \ref{modattack}, and consequently worked around by a suitable choice of parameters.  
\item A more rigorous complexity analysis is provided.
\item The scheme is extended in Sec. \ref{sec:extensions} to work over a reshaped database (viewed as a $t$-dimensional hypercube) in order to provide good rates when the files are small (with respect to the number of them) and the upload cost cannot be neglected. 
\item The new attack proposed in \cite{lage2025securitycodebasedpirscheme} and a way to circumvent it are addressed briefly in Remark \ref{newattack}. The paper will soon be updated to reflect the changes required to circumvent this attack.
\item Thorough comparisons to the closest rival schemes (XPIR, SimplePIR) are carried out in Sec. \ref{sec:comparison}, showing that our scheme compares favorably.
\end{itemize}

\paragraph{Notation:}
Throughout this paper $q$ is a prime power, and we denote a finite field of size $q$ by $\F_q$ and its multiplicative group by $\F_q^\times=\F_q \setminus \lbrace 0 \rbrace$. The extension field $\F_{q^s}$ can be seen as a vector space of dimension $s$ over $\F_q$. For a set of linearly independent vectors $\Gamma = \lbrace \gamma_1, \dots , \gamma_v\rbrace \subset \F_{q^s}$ we denote by $\langle \gamma_1, \dots , \gamma_v \rangle_{\F_q} \subset \F_{q^s}$ the vector subspace of dimension $v$ over $\F_q$. The corresponding projection map is denoted by $\psi_\Gamma : \F_{q^s} \to \langle \gamma_1, \dots , \gamma_v \rangle_{\F_q}$.

For a vector $x \in \F_q^t$ and an ordered set $J \subset [m]=\{1,\ldots,m\}$ of size $t$ we define $\phi_J: \F_q^t \to \F_q^m$  to be the extension of $x$ with zeroes at indices $j \not\in J$. \emph{E.g.,} for $J=\lbrace 1,3\rbrace$ and $m = 5$, $\phi_J([x_1,x_2]) = [x_1, 0, x_2,0,0]$. For a set $I \subseteq [n]$ we denote the complement of this set by $\bar I = [n] \setminus I$.

We parametrize a linear code over $\mathbb{F}_q$ by its length $n$, dimension $k$, and minimum Hamming distance $d$.  A linear $[n,k,d]_q$ code is capable of  correcting $d-1$ erasures or $t \leq  \floor{\frac{d-1}{2}}$ errors. We may omit  $q$ from the notation when clear from context or not directly important. The Hamming  weight of a vector $y$ is defined as the number of nonzero coordinates and denoted by $\wt(y)$.

\begin{table}[H]
\begin{tabular}{|l|l|}
\hline
$q$      & size of the base field $\F_q$ \\\hline
$s$      & degree of the extension field $\F_{q^s}$ over the base field \\\hline
$n$      & length of the code \\\hline
$k$      & dimension of the code \\\hline
$v$      & dimension of the subspace $V$ of $\F_{q^s}$ seen as an $\F_q$-linear vector space  \\\hline
$w = s-v$& dimension of the subspace $W$ of $\F_{q^s}$ seen as an $\F_q$-linear vector space \\\hline
$\delta \leq (n-k)(s-v)$ & number of columns in a file matrix (level of subpacketization)  \\\hline
$m$      & number of files stored on the database \\\hline
$L$      & number of rows in a file matrix \\\hline         
\end{tabular}
\caption{Important parameters used in CB-cPIR.}
\end{table}

The rest of the paper is organized as follows. In the remaining part of this introductory section, we will give a brief overview of code-based cryptography by introducing the error-decoding problem for a random linear code and the classic McEliece cryptosystem built on the known hardness of this problem. The basic model for computational PIR is also introduced. In Sec. \ref{sec:HHW}, we lay out the original HHW scheme and recall the observed weaknesses. Sec. \ref{sec:cb-cpir} then introduces the CB-cPIR scheme and demonstrates how it circumvents the identified attacks. Some modified and new attacks are exposed as well, which can also be avoided with appropriate parameter changes. In Sec. \ref{sec:comparison} we compare the new scheme to some state-of-the-art baseline works, and Sec. \ref{sec:conclusion} concludes the paper.

\subsection{Hardness of decoding a random linear code}
The security of the CB-cPIR scheme inherently relies on the assumption that decoding a random linear code is hard \cite{Berlekamp1978}. 

Let $G$ be an arbitrary, publicly accessible generator matrix for a random linear $[n,k,d]$  code $C \subset \F_{q}^n$. Then, given a secret message $x \in \F_{q}^k$ and a secret error vector $e \in \F_{q}^n$ of Hamming weight $\wt(e) = t\leq \floor{\frac{d-1}{2}}$, both chosen uniformly at random from their respective sample space, the decoding assumption asserts that for any random vector $r \in \F_{q}^n$ we have 
$$y=xG + e \compequi r, $$
where $\compequi$ denotes computational indistinguishability.
This assumption is utilized in a public-key cryptosystem presented by R.~J.~McEliece \cite{McEliece}, along with the assumption that a certain structured code is indistinguishable from a random linear code. In our PIR scheme we utilize a genuinely random linear code and the latter assumption will be redundant.

\begin{remark}
    We often omit the minimum distance $d$ when describing a random $[n,k,d]$ linear code and simply refer to it as an $[n,k]$ linear code. For sufficiently large $q$, a random linear code is MDS with high probability, making the minimum distance $d$ implicit.
\end{remark}

\paragraph{McEliece Cryptosystem:} 
The cryptosystem published by McEliece is based on Goppa codes \cite{Goppa}, which are well-known structured algebraic geometry codes. For binary Goppa codes, there is a fast decoding algorithm given by Patterson \cite{PattersonAlgo}. 

For the key generation, we construct a generator matrix $G$ for an $[n,k,d]$ Goppa code $C$, which can correct $t\leq \floor{\frac{d-1}{2}}$ errors. Then, we sample a random scrambling matrix $S$ and a permutation matrix $P$, and publish the public generator matrix $G' = SGP$ generating a (seemingly random) linear code with the same parameters as the code $C$. A user can then encrypt and transmit a message $x \in \F_q^k$ as $y = xG' + e$, where $e \in \F_q^n$ is an error vector randomly generated by the user with Hamming weight $\wt(e)=t$.

On receiving the transmission from the user, we compute $y' = yP^{-1} = xSG + eP^{-1}$. Noticing that $xSG \in C$ and $\wt(xP^{-1}) = t$, we can then efficiently decode $y'$ using Patterson's algorithm to obtain $x' = xS$ and invert to obtain the decrypted message $x = x'S^{-1}$. Guessing the generator matrix $G$ from the public generator matrix $G'$ is infeasible due to the astronomical number of choices for the scrambling matrix $S$ and permutation matrix $P$. However, if the Goppa polynomial and evaluation points are known or determined, then $P$ can be determined in polynomial time by the support splitting algorithm (SSA) \cite{SSA}. 

In the context of the CB-cPIR scheme, this is irrelevant since the code can actually be randomly chosen (as the user only needs to perform erasure decoding). This is in contrast to the structured Goppa code disguised by the scrambling and permutation matrices. The best known approach for an attack involves correctly guessing an information set of the given code. This probabilistic decoding method is described in its most naïve form by Prange's algorithm \cite{PrangeISD}. The runtime for this algorithm is exponential given error vectors with a suitably chosen weight, which will be the basis of our security assumptions. We give concrete values of the parameters used in Section \ref{parameters}.

\subsection{Computational private information retrieval}
\label{ssPIR}
\emph{Private information retrieval} is the process of downloading a file from a database without revealing to the database the identity of the desired file.

\paragraph{Database:}
In the computational setting the considered database is a single server consisting of $m$ files, which we will represent by a matrix $X \in \F_q^{L \times m\delta}$. Each file in this database is represented by a submatrix $X^j \in \F_q^{L \times \delta}$, where the parameter $L$ describes the size of the file and $\delta \leq (n-k)(s-v)$ can be considered as the level of subpacketization required by the scheme.

\begin{figure}[H]
  \centering
  \def\x{*1}

\begin{tikzpicture}

  \coordinate (Xnw) at (0\x,1\x);
  \node[draw=none] () at ($(Xnw)+(-0.5\x,-0.4\x)$) {$X = $};

  \draw[draw=black] (Xnw) rectangle ($(Xnw)+(1.2\x,-0.8\x)$) node[pos=0.5] {$X^1$};
  \draw[draw=black] ($(Xnw)+(1.2\x,0\x)$) rectangle ++(1.2\x,-0.8\x) node[pos=0.5] {$X^2$};
  \draw[draw=black] ($(Xnw)+(2.4\x,0\x)$) rectangle ++(1.2\x,-0.8\x) node[pos=0.5] {$X^3$};
  \node[draw=none] at ($(Xnw)+(4.2\x,-0.4\x)$) {$\cdots$};
  \draw[draw=black] ($(Xnw)+(4.8\x,0\x)$) rectangle ++(1.2\x,-0.8\x) node[pos=0.5] {$X^m$};
    
  \draw [decorate,decoration={brace,amplitude=3pt}] ($(Xnw)+(6.1\x,0\x)$) -- ++(0\x,-0.8\x) node [black,midway,xshift=0.3\x cm] {$L$};
  \draw [decorate,decoration={brace,amplitude=3pt}] ($(Xnw)+(1.2\x,0.1\x)$) -- ++(1.2\x,0\x) node [black,midway,yshift=0.4\x cm] {$\delta$};
  
\end{tikzpicture}

  \caption{Illustration of the file matrix $X$.}
  \label{fig:fileMatrix}
\end{figure}

\begin{definition}
    Consider a database $X \in \F_q^{L \times m\delta}$ of $m$ files that are stored on a single server as described above. A computational PIR scheme for such a storage system consists of the following:
    \begin{itemize}
        \item Queries $(X, \mathcal{S}, \mathcal{P}, i) \mapsto Q^i$: For a given index $i \in [m]$ generate a query $Q^i$ from a set of secret information $\mathcal{S}$  and a set of public information $\mathcal{P}$. 
        \item Response $(X, Q^i) \mapsto A^i$: Given a query $Q^i$, the server computes an answer given by $A^i = X\cdot Q^i$ and transmits it to the user.
        \item Data reconstruction $(A^i, \mathcal{S}, \mathcal{P}) \mapsto X^i$: A function which takes as an argument the server answer $A^i$ and returns the desired file $X^i$.
    \end{itemize}
\emph{Correctness:} A PIR scheme is said to be \emph{correct} if the user can successfully recover the desired file from the server response.
\\\\
\emph{Security:} The database should not be able to deduce any information about the desired file index from the user's query chosen from the set of all possible queries $Q$. A PIR scheme is said to be $(T, \epsilon)\emph{-secure}$ if for any computationally constrained adversarial algorithm $\mathcal{A}: (X, Q) \to [m]$ running in time at most $T$, and for any $i,j \in [m]$ we have
$$|\prob[\mathcal{A}(X,Q^i) = i] - \prob[\mathcal{A}(X,Q^j) = i]| \leq \epsilon.$$
That is, any adversary running in time $T$ can distinguish between any two queries $Q^i$ and $Q^j$ with advantage at most $\epsilon$. 

\end{definition}

The rate of a PIR scheme measures its efficiency as the ratio between the size (denoted by $|\cdot|$) of the desired file and the total cost of communication. For the protocol to be nontrivial the rate must be greater than $1/m$. That is, it must be more efficient than trivially downloading the entire database. 

\begin{definition}
    The total \emph{communication complexity} is defined as 
    $$C_{\mathrm{total}} = \textit{upload cost} + \textit{download cost} = |Q^i| + |A^i|.$$
\end{definition}
\begin{definition}
\label{ratedef}
    The \emph{rate} of a PIR scheme is defined as
    $$R_{\pir} = \frac{\textit{size of desired file}}{C_{\mathrm{total}}} = \frac{|X^i|}{|Q^i| + |A^i|}.$$
\end{definition}

\section{Outline of the original HHW PIR scheme}
\label{sec:HHW}
In this section we describe the first code-based computational PIR scheme \cite{holzbaur2020isit}, which we will henceforth refer to as the HHW scheme. The HHW scheme made use of the assumption that decoding a random linear code is hard, to query a database consisting of a single server. The queries are cleverly constructed with a backdrop of codewords from a random linear code and specifically crafted errors, which will allow the user to efficiently decode the servers response and correctly reconstruct the file they desire.

The HHW scheme was shown to be vulnerable to a distinguishability attack \cite{bordage2020privacy} due to discernible rank differences in submatrices of the query. In Section \ref{SecurityofCBC} we circumvent this subquery attack and fix the HHW PIR scheme. We first we outline the HHW scheme, which will be the basis of the rest of the paper. 

\subsection{System model}
We are concerned with a single-server data storage containing $m$ files of size $L \times \delta$ over $\F_{q}$, where $\delta \leq (n-k)(s-v)$ and the parameters $n,k,s,v$ are as specified below. The data content on this server is denoted by $X \in \F_q^{L \times m\delta}$ as specified in Section \ref{ssPIR}.

\paragraph{Queries:}
To construct the queries, the user samples a set of public information $\mathcal{P}$ and a set of secret information $\mathcal{S}$ as follows:

The public information $\mathcal{P} = \lbrace G_C \rbrace$ consists of a generator matrix $G_C$ of a random linear code $C \subset \F_{q^s}^n$ of dimension $k$ sampled uniformly at random from the set of all possible $[n,k]_{q^s}$ linear codes.

Having selected the code $C$ the user samples uniformly at random the following secret information $\mathcal{S} = \lbrace I, D, \Gamma, V, W, E, \Delta \rbrace$:

\begin{itemize}
    \item An information set $I$ of $C$ with $|I|=k$.
    \item A matrix $D \in \F_{q^s}^{m\delta \times n}$ such that each row of $D$ is a codeword in $C$.
    \item A basis $\Gamma = \lbrace \gamma_1, \dots , \gamma_s\rbrace$ of $\F_{q^s}$ over $\F_{q}$, and the vector subspaces $V = \langle \gamma_1, \dots , \gamma_v \rangle_{\F_q}$ and $W = \langle \gamma_{v+1}, \dots , \gamma_s \rangle_{\F_q}$. 
    \item A matrix $E_0 \in V^{m\delta \times (n-k)}$.
    \item A \emph{full rank} matrix $\Delta_0 \in W^{\delta \times (n-k)}$.
\end{itemize}
We then expand the matrices $E_0$ and $\Delta_0$ such that their column support lies off of the chosen information set $I$. That is we expand to the matrices $E$ and $\Delta$ given by $E = \phi_{\bar I}(E_0)\in V^{m\delta \times n}$, and the full-rank matrix $\Delta = \phi_{\bar I}(\Delta_0)\in W^{\delta \times n}$.

Finally, for any desired file index $i \in [m]$ the user constructs the query $(X, \mathcal{S}, \mathcal{P}, i) \mapsto Q^i$ as $$Q^i = D + E + e_i^m \otimes \Delta.$$ Where $e_i^m \in \F_{q^s}^m$ is the $i^{th}$ standard basis vector and $\otimes$ is the matrix Kronecker product. An illustration of the query matrix is given in Fig. \ref{fig2}. 

\begin{figure}[H]
\centerline{\def\x{*1}

\begin{tikzpicture}
  
  \coordinate (Dnw) at (0\x,2\x);
  \node[draw=none] () at ($(Dnw)+(-0.5\x,-2\x)$) {$Q^i = $};

  \draw[draw=black] (Dnw) rectangle ($(Dnw)+(2\x,-4\x)$) node[pos=0.5] {$D$};
  \node[draw=none] at ($(Dnw)+(2.3\x,-2\x)$) {$+$};

  \draw [decorate,decoration={brace,amplitude=3pt}] ($(Dnw)+(0\x,0.1\x)$) -- ++(2\x,0\x) node [black,midway,yshift=0.3\x cm] {$n$};

  \coordinate (Enw) at ($(Dnw) + (2.6\x,0\x)$);

  \node[draw=none] at ($(Enw)+(2.3\x,-2\x)$) {$+$};

  \draw[draw=none, fill = blue!100!white] ($(Enw)+(0.0\x,0\x)$) rectangle ++(0.3\x,-4\x);
  \draw[draw=none, fill = blue!100!white] ($(Enw)+(0.55\x,0\x)$) rectangle ++(0.3\x,-4\x);
  \draw[draw=none, fill = blue!100!white] ($(Enw)+(1.3\x,0\x)$) rectangle ++(0.3\x,-4\x);

  \draw[draw=black] (Enw) rectangle ($(Enw)+(2\x,-4\x)$) node[pos=0.5] {$E$};
  
  \coordinate (Knw) at ($(Enw) + (2.6\x,0\x)$);

  \draw [decorate,decoration={brace,amplitude=3pt}] ($(Knw)+(2.1\x,0\x)$) -- ++(0\x,-4\x) node [black,midway,xshift=0.4\x cm, rotate=270] {$m\delta$};

  \coordinate (Deltanw) at ($(Knw) + (0\x,-2.4\x)$);

  \draw[draw=none, fill=green!100!white, dotted] ($(Deltanw)+(0\x,0\x)$) rectangle ++(0.3\x,-0.8\x);
  \draw[draw=none, fill=green!100!white,dotted] ($(Deltanw)+(0.55\x,0\x)$) rectangle ++(0.3\x,-0.8\x);
  \draw[draw=none, fill=green!100!white,,dotted] ($(Deltanw)+(1.3\x,0\x)$) rectangle ++(0.3\x,-0.8\x);

  \draw[draw=black, dashed] (Deltanw) rectangle ($(Deltanw)+(2\x,-0.8\x)$) node[pos=0.5] {$\Delta$};

  \draw[draw=black] (Knw) rectangle ($(Knw)+(2\x,-4\x)$) node[pos=0.5] {$e_i^m \otimes \Delta $};

\end{tikzpicture}

}
\caption{Illustration of the query matrix $Q^{i}$.}
\label{fig2}
\end{figure}

\paragraph{Retrieval:}
Decompose $Q^i$ as the stack of of submatrices $Q^i_1, \dots Q^i_m \in \F_{q^s}^{\delta \times n}$. The server upon receiving the query responds with
\begin{align*}
    A^i & = X\cdot Q^i 
     = \begin{bmatrix}
         X^1 & \cdots & X^m
     \end{bmatrix} 
     \begin{bmatrix}
         Q^i_1 \\
         \vdots \\
         Q^i_m
     \end{bmatrix}
     = \sum_{j=1}^m X^j\cdot Q^i_j \\
     & = \sum_{j=1}^m X^j\cdot D_j + \sum_{j=1}^m X^j\cdot E_j  + X^i\cdot \Delta.
\end{align*}
The rows of the matrix $\sum_{j=1}^n X^j\cdot D_j$ lie in $C$ and the rows of $\sum_{j=1}^n X^j\cdot E_j  + X^i\cdot \Delta$ have support $\bar I$ . Therefore by erasure decoding we can obtain $B^i = \sum_{j=1}^n X^j\cdot E_j  + X^i\cdot \Delta$.
 We can then project onto the space W and get $\psi_W(B^i) = X^i\cdot \Delta$. Since $\Delta$ has full rank we can recover the desired file $X^i$.
 
\subsection{Rate of the scheme} Let us next recall the rate achievable by the HHW scheme. 

The size of the desired file in bits is $|X^i| = \delta L \log(q)$. The size of the query uploaded by the user is $|Q^i| = m\delta n\log(q^s)$. The size of the answer provided by the server, \emph{i.e.}, the download cost for the user is $|A^i| = Ln \log(q^s)$. This with definition \ref{ratedef} gives us the rate of the HHW scheme.

\begin{theorem}\cite[Thm~1]{holzbaur2020isit} The rate of the HHW scheme is 
$$R_{\pir} = \frac{L\delta \log(q)}{(m\delta n + Ln) \log(q^s)}=\frac{L\delta }{ns(m\delta  + L)}.$$
\end{theorem}

\begin{corollary}\label{approxRate1} \cite[Cor.~1]{holzbaur2020isit} Assume $L>>\delta m$, \emph{i.e.}, the size of the files is large compared to the number of them and we can safely ignore the upload cost. Then the rate of the scheme is 
\begin{align*}
    R_{\pir} & \approx \frac{\delta}{ns}
     \leq 1 - \frac{k + \frac{v}{s}(n-k)}{n}.
\end{align*}

\end{corollary}

\subsection{Security}

\paragraph{Information set decoding:}
One obvious way to attack the HHW scheme is by information set-decoding the query. Let $G$ be the public generator matrix for the code $[n,k]_{q^s}$ linear code $C$. Then the information set decoding attack involves guessing the secret information set $I$ and inverting the full rank matrix $G_I$, which is $G$ restricted to the columns indexed by $I$. After guessing an information set the attacker can perform the operation $Q^i_I \cdot G_I^{-1} \cdot G$ to obtain the secret matrix $D$. The number of guesses required for the attacker to succeed is $\binom{n}{k}$ which -including the cost of matrix inversion- ultimately gives us the work factor $$\text{Wf} = k^3 \binom{n}{k}.$$
\begin{remark}
    After decoding the query matrix the attacker must additionally distinguish between errors from the different subspaces $V$ and $W$. This can be done in polynomial time.
\end{remark}  
In \cite{holzbaur2020isit} the authors suggest the parameters $n=100, k=50$ for which the work factor is $\text{Wf} = 50^3 \binom{100}{50} \approx 2^{113}$. These parameters therefore offer $113$ bit security in the context of the information set decoding attack.

However, there might be other forms of attacks, some of which were identified and shown to have infeasible complexities in \cite{holzbaur2020isit}.

\paragraph{Subquery attack:}

Despite being resistant to several forms of attack the HHW scheme was shown to be insecure to a specific form of distinguishability attack due to discernible rank differences in submatrices of the query.

We describe this subquery attack found in \cite{bordage2020privacy}.  Consider the submatrices $Q^i[j]$ of the received query where the rows $[(j-1)\delta +1,j\delta]$ of $Q^i$ are deleted, $j \in [m]$. It was shown in \cite{bordage2020privacy} that we can decompose $$\F_{q^s}^n = C ~ \oplus ~ \phi_{\bar I}(V^{n-k}) ~ \oplus  ~ \phi_{\bar I}(W^{n-k}).$$
Due to this fact the $\F_q$-rank of a submatrix $$\rank(Q^i[j]) = \rank(D[j]+E[j]) + \rank(e^i[j] \otimes \Delta).$$ 
For $j \neq i$, $\rank(Q^i[j]) = \rank(D[j]+E[j]) + \delta$, and $\rank(Q^i[i]) = \rank(D[i]+E[i]) \leq ns - \delta$.

The attack involves computing the $\F_q$-rank of all $m$ submatrices $Q^i[j]$ and discerning the desired file index due to the low rank of $Q^i[i]$. Discerning the desired file index is only possible if $\rank(Q^i[i]) < \rank(Q^i[j])$ for all $j \neq i$, that is,  the attack fails if $\rank(D[j] + E[j]) < ns - 2\delta$. In \cite{bordage2020privacy} the authors prove that the probability 
\begin{align*}
    p ~ & := \prob(\rank(D[j] + E[j]) < ns - 2\delta) \\
    & \leq \binom{ns - \delta}{ns-2\delta}_q q^{-\delta^2(m-1)} 
     \leq q^{(\delta + 1)(ns - 2\delta) - \delta^2(m-1)}.
\end{align*}
As long as $(\delta + 1)(ns - 2\delta) < \delta^2(m-1)$ this probability is meaningful. 
In other words, when $m > 1 + \frac{(\delta + 1)(ns - 2\delta)}{\delta^2}$ the attack can discern the desired file index with high probability, thereby breaking the scheme for an unbounded number of files.

\begin{theorem}\cite[Thm~3.4]{bordage2020privacy}
For a given database $X$ containing $m > 1 + \frac{(\delta + 1)(ns - 2\delta)}{\delta^2}$ files, there exists an algorithm $\mathcal{A}: (X,Q) \to [m]$ running in time $\mathcal{O}(m^2(ns)^3)$ which can recover the desired file index $i$ from a query $Q^i$ constructed as per the HHW PIR scheme with probability at least $$1-q^{(\delta + 1)(ns - 2\delta) - \delta^2(m-1)},$$
where the probability is taken over the randomness of the query generation.
\end{theorem}

\begin{corollary}
    The HHW scheme is not $(T,\epsilon)$-secure against an adversary running in time $T \geq \mathcal{O}(m^2(ns)^3)$.
\end{corollary}

\section{The CB-cPIR scheme}
\label{sec:cb-cpir}
Let us now introduce the CB-cPIR scheme, which is a modification of the original HHW scheme. In the original scheme \cite{holzbaur2020isit} the secret in the query came from the standard unit vector $e_i^m$. The attack in \cite{bordage2020privacy} with high probability can reveal this secret due to the fact that the standard unit vector has low weight. 

In the CB-cPIR scheme a query consists of a concatenation of two independent queries constructed as prescribed by the HHW scheme, with the key difference now being that the secrets will be of high weight. This prevents the submatrices of the query from having discernible rank differences and allows us to circumvent the subquery attack. 

The database setup for this scheme remains exactly the same as in the HHW scheme. The CB-cPIR protocol is described algorithmically in Fig. \ref{CBCpirAlgo}.

\subsection{System model}

\paragraph{Queries:}
The user will construct two independent queries as prescribed by the HHW scheme.

As in the HHW scheme for each individual query $Q_j$, where $j \in \lbrace 1,2 \rbrace$, the user samples independently and uniformly at random a set of public information and a set of secret information. That is, for each query we have a set of public information $\mathcal{P}_j = \lbrace G_{C_j} \rbrace$ and secret information $\mathcal{S}_j = \lbrace I_j, D_j, \Gamma_j, V_j, W_j, E_j, \Delta_j \rbrace$. The complete set of public information is then $\mathcal{P} = \mathcal{P}_1 \cup \mathcal{P}_2$. And the complete set of secret information is $\mathcal{S}= \mathcal{S}_1 \cup \mathcal{S}_2 \cup \lbrace \beta \rbrace$. Where $\beta \in {\F_q^\times}^m$ is a vector of full weight sampled uniformly at random by the user.

For any desired file index $i \in [m]$, the user then constructs the queries $$Q_1 = D_1 + E_1 + v_1 \otimes \Delta_1 ~ \text{and} ~ Q_2 = D_2 + E_2 + v_2 \otimes \Delta_2,$$
where $v_1 = \beta$ and $v_2 = \beta + e^m_i$.

\begin{figure}[htbp]
\centerline{\def\x{*1}

\begin{tikzpicture}
  
  \coordinate (Dnw) at (0\x,2\x);
  \node[draw=none] () at ($(Dnw)+(-0.5\x,-2\x)$) {$Q_1 = $};

  \draw[draw=black] (Dnw) rectangle ($(Dnw)+(2\x,-4\x)$) node[pos=0.5] {$D$};
  \node[draw=none] at ($(Dnw)+(2.3\x,-2\x)$) {$+$};

  \draw [decorate,decoration={brace,amplitude=3pt}] ($(Dnw)+(0\x,0.1\x)$) -- ++(2\x,0\x) node [black,midway,yshift=0.3\x cm] {$n$};

  \coordinate (Enw) at ($(Dnw) + (2.6\x,0\x)$);

  \node[draw=none] at ($(Enw)+(2.3\x,-2\x)$) {$+$};

  \draw[draw=none, fill = blue!100!white] ($(Enw)+(0.0\x,0\x)$) rectangle ++(0.3\x,-4\x);
  \draw[draw=none, fill = blue!100!white] ($(Enw)+(0.55\x,0\x)$) rectangle ++(0.3\x,-4\x);
  \draw[draw=none, fill = blue!100!white] ($(Enw)+(1.3\x,0\x)$) rectangle ++(0.3\x,-4\x);

  \draw[draw=black] (Enw) rectangle ($(Enw)+(2\x,-4\x)$) node[pos=0.5] {$E$};
  
  \coordinate (Knw) at ($(Enw) + (2.6\x,0\x)$);

  \draw [decorate,decoration={brace,amplitude=3pt}] ($(Knw)+(2.1\x,0\x)$) -- ++(0\x,-4\x) node [black,midway,xshift=0.4\x cm, rotate=270] {$m\delta$};

  \coordinate (Deltanw) at ($(Knw) + (0\x,-2.4\x)$);

  \draw[draw=none, fill=green!100!white, dotted] ($(Deltanw)+(0\x,0\x)$) rectangle ++(0.3\x,-0.4\x);
  \draw[draw=none, fill=green!100!white,dotted] ($(Deltanw)+(0.55\x,0\x)$) rectangle ++(0.3\x,-0.4\x);
  \draw[draw=none, fill=green!100!white,,dotted] ($(Deltanw)+(1.3\x,0\x)$) rectangle ++(0.3\x,-0.4\x);

  \draw[draw=black, dashed] (Deltanw) rectangle ($(Deltanw)+(2\x,-0.4\x)$) node[pos=0.5] {$\beta_{m-3}\Delta$};

  \draw[draw=black] (Knw) rectangle ($(Knw)+(2\x,-4\x)$) node[pos=0.5] {$\beta \otimes \Delta$};

  \coordinate (Deltanw) at ($(Knw) + (0\x,-2.8\x)$);

  \draw[draw=none, fill=green!100!white, dotted] ($(Deltanw)+(0\x,0\x)$) rectangle ++(0.3\x,-0.4\x);
  \draw[draw=none, fill=green!100!white,dotted] ($(Deltanw)+(0.55\x,0\x)$) rectangle ++(0.3\x,-0.4\x);
  \draw[draw=none, fill=green!100!white,,dotted] ($(Deltanw)+(1.3\x,0\x)$) rectangle ++(0.3\x,-0.4\x);

  \draw[draw=black, dashed] (Deltanw) rectangle ($(Deltanw)+(2\x,-0.4\x)$) node[pos=0.5] {$\beta_{m-2} \Delta$};

  \coordinate (Deltanw) at ($(Knw) + (0\x,-3.2\x)$);

  \draw[draw=none, fill=green!100!white, dotted] ($(Deltanw)+(0\x,0\x)$) rectangle ++(0.3\x,-0.4\x);
  \draw[draw=none, fill=green!100!white,dotted] ($(Deltanw)+(0.55\x,0\x)$) rectangle ++(0.3\x,-0.4\x);
  \draw[draw=none, fill=green!100!white,,dotted] ($(Deltanw)+(1.3\x,0\x)$) rectangle ++(0.3\x,-0.4\x);

  \draw[draw=black, dashed] (Deltanw) rectangle ($(Deltanw)+(2\x,-0.4\x)$) node[pos=0.5] {$\beta_{m-1}\Delta$};

  \coordinate (Deltanw) at ($(Knw) + (0\x,-3.6\x)$);

  \draw[draw=none, fill=green!100!white, dotted] ($(Deltanw)+(0\x,0\x)$) rectangle ++(0.3\x,-0.4\x);
  \draw[draw=none, fill=green!100!white,dotted] ($(Deltanw)+(0.55\x,0\x)$) rectangle ++(0.3\x,-0.4\x);
  \draw[draw=none, fill=green!100!white,,dotted] ($(Deltanw)+(1.3\x,0\x)$) rectangle ++(0.3\x,-0.4\x);

  \draw[draw=black, dashed] (Deltanw) rectangle ($(Deltanw)+(2\x,-0.4\x)$) node[pos=0.5] {$\beta_{m}\Delta$};

  \coordinate (Deltanw) at ($(Knw) + (0\x,-0\x)$);

  \draw[draw=none, fill=green!100!white, dotted] ($(Deltanw)+(0\x,0\x)$) rectangle ++(0.3\x,-0.4\x);
  \draw[draw=none, fill=green!100!white,dotted] ($(Deltanw)+(0.55\x,0\x)$) rectangle ++(0.3\x,-0.4\x);
  \draw[draw=none, fill=green!100!white,,dotted] ($(Deltanw)+(1.3\x,0\x)$) rectangle ++(0.3\x,-0.4\x);

  \draw[draw=black, dashed] (Deltanw) rectangle ($(Deltanw)+(2\x,-0.4\x)$) node[pos=0.5] {$\beta_{1}\Delta$};

  \coordinate (Deltanw) at ($(Knw) + (0\x,-0.4\x)$);

  \draw[draw=none, fill=green!100!white, dotted] ($(Deltanw)+(0\x,0\x)$) rectangle ++(0.3\x,-0.4\x);
  \draw[draw=none, fill=green!100!white,dotted] ($(Deltanw)+(0.55\x,0\x)$) rectangle ++(0.3\x,-0.4\x);
  \draw[draw=none, fill=green!100!white,,dotted] ($(Deltanw)+(1.3\x,0\x)$) rectangle ++(0.3\x,-0.4\x);

  \draw[draw=black, dashed] (Deltanw) rectangle ($(Deltanw)+(2\x,-0.4\x)$) node[pos=0.5] {$\beta_{2}\Delta$};
  \coordinate (Deltanw) at ($(Knw) + (0\x,-0.8\x)$);

  \draw[draw=none, fill=green!100!white, dotted] ($(Deltanw)+(0\x,0\x)$) rectangle ++(0.3\x,-0.4\x);
  \draw[draw=none, fill=green!100!white,dotted] ($(Deltanw)+(0.55\x,0\x)$) rectangle ++(0.3\x,-0.4\x);
  \draw[draw=none, fill=green!100!white,,dotted] ($(Deltanw)+(1.3\x,0\x)$) rectangle ++(0.3\x,-0.4\x);

  \draw[draw=black, dashed] (Deltanw) rectangle ($(Deltanw)+(2\x,-0.4\x)$) node[pos=0.5] {$\beta_3\Delta$};

  \coordinate (Deltanw) at ($(Knw) + (0\x,-1.2\x)$);

  \draw[draw=none, fill=green!100!white, dotted] ($(Deltanw)+(0\x,0\x)$) rectangle ++(0.3\x,-0.4\x);
  \draw[draw=none, fill=green!100!white,dotted] ($(Deltanw)+(0.55\x,0\x)$) rectangle ++(0.3\x,-0.4\x);
  \draw[draw=none, fill=green!100!white,,dotted] ($(Deltanw)+(1.3\x,0\x)$) rectangle ++(0.3\x,-0.4\x);

  \draw[draw=black, dashed] (Deltanw) rectangle ($(Deltanw)+(2\x,-0.4\x)$) node[pos=0.5] {$\beta_{4}\Delta$};

\end{tikzpicture}}
\caption{Illustration of the query matrix $Q_1$.}
\label{fig3}
\end{figure}
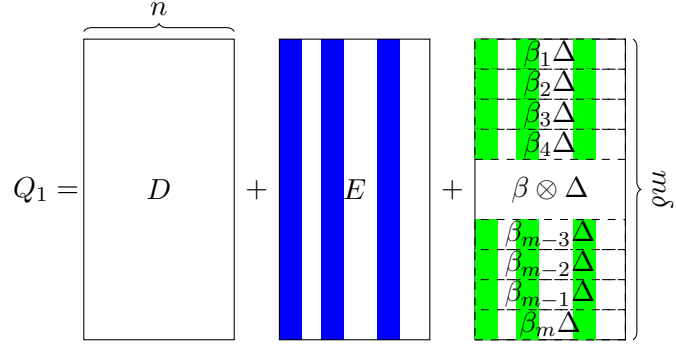

Finally the user concatenates these two queries and sends to the server the final query $$Q^i = [Q_1 | Q_2].$$
An illustration of the first query is given in Fig. \ref{fig3}.

\paragraph{Retrieval:}
The server upon receiving the query responds with $$A^i  = X\cdot Q^{i} = X\cdot [Q_1 | Q_2] = [X\cdot Q_1 | X \cdot Q_2] = [A_1 | A_2].$$
For each $j \in \lbrace 1,2 \rbrace$ we decompose $Q_j$ as the stack of the submatrices $Q_j^1, \dots, Q_j^m \in \F_{q^s}^{\delta \times n}$. We then have 
\begin{align*}
    A_j & = X\cdot Q_j \\
    & = \begin{bmatrix}
         X^1 & \cdots & X^m
     \end{bmatrix} 
     \begin{bmatrix}
         Q_j^1 \\
         \vdots \\
         Q_j^m
     \end{bmatrix} \\
    & = \sum_{k=1}^m X^k\cdot Q_j^k \\
    & = \sum_{k=1}^m X^k\cdot D_j^k + \sum_{k=1}^m X^k\cdot E_j^k  +  X \cdot  (v_j \otimes \Delta_j).
\end{align*}
The rows of the matrix $\sum_{k=1}^n X^k\cdot D_j^k$ lie in $C_j$ and the rows of $\sum_{k=1}^n X^k \cdot E_j^k  + X \cdot (v_j \otimes \Delta_j)$ have support $\bar I$ . 
Therefore by erasure decoding we can obtain $$B_j = \sum_{k=1}^m X^k\cdot E_j^k  + X \cdot (v_j \otimes \Delta_j).$$

We can then project onto the space W and get $\psi_W(B_j) = X \cdot (v_j \otimes \Delta_j)$. Since by construction $\Delta_j$ has full rank we can recover $ R_j = X \cdot (I_{\delta \times \delta} \otimes v_j) $. 
Finally the user can retrieve their desired file
\begin{align*}
    R &= R_2 - R_1 \\
    &= X \cdot (I_{\delta \times \delta} \otimes (v_2 - v_1))\\
    &= X \cdot (I_{\delta \times \delta} \otimes e^m_i)\\ 
    &= X^i.
\end{align*}

\begin{algorithm}[H]
\scriptsize
\caption*{\textbf{Construct cB-cPIR:} the parameters used are $n,k,s,v$ which decide the security of the protocol. The database consists of $m$ file matrices in $\F_q^{L\times\delta}$ represented as a matrix $X \in \F_q^{L \times m\delta}$, where $\delta := (n-k)(s-v)$ is the required level of subpacketization.}

\begin{algorithmic}
\State \textbf{SecretivelySample}$(n,k,s,v) \to \mathcal{S}$.
\begin{itemize}
    \item Sample $C \xleftarrow{\$} Gr_k(\F_{q^s}^n)$. \Comment{$[n,k]_{q^s}$ random linear code}
    \item Sample $D \xleftarrow{\$} C^{m\delta \times 1}$. \Comment{matrix of random codewords}
    \item Sample $I \xleftarrow{\$} \binom{[n]}{k}$. \Comment{random information set}
    \item Sample $\Gamma = \lbrace \gamma_1, \dots , \gamma_s\rbrace\xleftarrow{\$} \mathcal{B}_{\F_q}(\F_{q^s})$.  \Comment{random basis of $\F_{q^s}$ over $\F_q$}
    
    Split $\Gamma$ and generate subspaces $V,W = \langle \gamma_1, \dots , \gamma_v \rangle_{\F_q}, \langle \gamma_{v+1}, \dots , \gamma_s \rangle_{\F_q}$.
    \item Sample $E_0 \xleftarrow{\$} V^{m\delta \times k}$. \Comment{masking error matrix}

    Generate $E \leftarrow \phi_{\bar I}(E_0)\in V^{m\delta \times n}$.
    \item Sample $\Delta_0 \xleftarrow{\$} \lbrace M \in W^{\delta \times k} ~|~\rank_{\F_q}(M) = \min(\delta,k)\rbrace$. \Comment{desired error matrix}

    Generate $\Delta \leftarrow \phi_{\bar I}(\Delta_0)\in W^{\delta \times n}$.
    \item Return $\mathcal{S} = \lbrace C, I, D, \Gamma, V, W, E, \Delta \rbrace$.
\end{itemize}

\State \textbf{Query}$(i\in [m]) \to Q$.
\begin{itemize}
    \item $S_1 \leftarrow \textbf{SecretivelySample}(n,k,s,v)$

    $S_2 \leftarrow \textbf{SecretivelySample}(n,k,s,v)$
    \item Sample $\beta \xleftarrow{\$} \F_q^m$. 
    \item Generate $Q_1 \leftarrow (D_1 + E_1 + \beta \otimes \Delta_1) \in \F_{q^s}^{m\delta \times n}$.

    Generate $Q_2 \leftarrow (D_2 + E_2 + (\beta + e_i^m) \otimes \Delta_2) \in \F_{q^s}^{m\delta \times n}$.
    \item Concatenate $Q = [Q_1 | Q_2] \in \F_{q^s}^{m\delta \times 2 n}$.
    \item Return $Q$.
\end{itemize}
\State \textbf{Answer}$(Q \in \F_{q^s}^{m\delta\times n}, X \in \F_q^{L \times m\delta }) \to A$.
\begin{itemize}
    \item Return $A = [A_1 | A_2] \leftarrow X\cdot Q \in \F_{q^s}^{L \times 2n}$.
\end{itemize}

\State \textbf{Recover}$(A \in \F_{q^s}^{L \times n}, \mathcal{S}_1,\mathcal{S}_2, i) \to R$
\begin{itemize}
    \item $\text{Err}_1 \leftarrow \text{erasureDecode}_{C_1,I_1}(A_1)$.
    \item $\text{Err}_2 \leftarrow \text{erasureDecode}_{C_2,I_2}(A_2)$.
    \item $R_1 \leftarrow \Delta_1^{-1}\cdot\psi_{W_1}(Err_1)$.
    \item $R_2 \leftarrow \Delta_2^{-1}\cdot\psi_{W_2}(Err_2)$.
    \item Return $R \leftarrow R_2 - R_1\in \F_q^{L \times \delta}$.
\end{itemize}
\label{CBCpirAlgo}
\end{algorithmic}

\end{algorithm}

\subsection{Rate of the CB-cPIR scheme}
Let us now look into the CB-cPIR scheme in more detail. The size (in bits) of each file in the database is $L\delta \log_2(q)$. The size of each query $Q^i$ is $2m\delta n \log_2(q^s)$ with response size $2Ln\log_2(q^s)$.

\begin{theorem}
  The PIR rate of the scheme is $$R_{\pir} = \frac{L\delta}{2(m\delta + L)ns}.$$
\end{theorem}

\begin{corollary}\label{approxRate}
    Assume $L>> m \delta $, \emph{i.e.}, the size of the files is large compared to the number of them and we can safely ignore the upload cost. Then the rate of the scheme is,
    \begin{align*}
      R_{\pir} & \approx \frac{\delta}{2ns}
       = \frac{1}{2}\left(1 - \frac{k + \frac{v}{s}(n-k)}{n}\right) .
    \end{align*}
\end{corollary}

\begin{corollary}
    The decoded response $R_1$ from the query $Q_1$ if stored can be reused for subsequent private file retrievals. The amortized rate for $f$ private file retrievals will then be $$R_{\pir} = \frac{f\delta}{(f+1)ns},$$ which for an increasing number of files $f$ approaches 
    $$\lim_{f\rightarrow \infty}R_{\pir} = \frac{\delta}{ns}$$ 
    matching the rate of the HHW scheme as given in Corollary \ref{approxRate1}.
\end{corollary} 

\subsection{Server and user complexity}

In this section, we concretely determine the total computational costs incurred by the server and the user in the private retrieval of a single file from the database. 

\paragraph{Server complexity:}
The server on receiving a query $Q^i \in \F_{q^s}^{m\delta \times 2n}$ responds with $A^i = X\cdot Q^i$. The concrete cost of naively multiplying these matrices is $2Lm\delta ns$ multiplications in $\F_q$.

\paragraph{User complexity:}
The user complexity for the PIR protocol is divided into two parts, the complexity of generating the query and the complexity of decoding the server response.
\begin{itemize}
    \item Query generation: The complexity of query generation is dominated by the following steps in the protocol: \begin{itemize}
        \item Generating the random codeword matrix $D$: This requires $m\delta kns$ multiplications over $\F_q$. 
        \item Generating the $V$ noise matrix $E$: This requires $m\delta kvs$ multiplications over $\F_q$.
        \item Generating the $W$ noise matrix $\Delta$: This requires $\delta kws$ multiplications over $\F_q$.     
        \item Kronecker product: This requires $\min(m,q)\delta k s$ multiplications over $\F_q$. The term $\min(m,q)$ arises from the fact that when $q > m$ there is a pigeon-holing of scalar multiplications in the Kronecker product. 
    \end{itemize}
    The above computational costs are incurred twice, once for each part of the query matrix. The total complexity for query generation is then:
    $$C_{Qgen} = 2\delta ks(mn + mv + w + \min(m,q)).$$
    \item Decoding response: The decoding of the response is dominated by the following steps in the protocol:
    \begin{itemize}
        \item Erasure decoding the response: This involves $Lkns$ multiplications over $\F_q$.
        \item Projection onto the subspace $W$: This involves viewing each $\F_{q^s}$ element as a vector representation in terms of our chosen secret basis $\Gamma$. This requires $Lks^2 + Lkw$ multiplications over $\F_q$.
        \item Inverting the matrix $\Delta$: This involves $Lk\delta s$ multiplications over $\F_q$.
    \end{itemize}
    The above computational costs are incurred twice, once for each part of the response matrix. The total complexity to decode the response is then:
    $$C_{Adec} = 2Lk(ns + s^2 + w + \delta s).$$
\end{itemize}

\begin{example}
Suppose we want to privately retrieve a single file with index $i$ from the server with elements in $\F_3$.  We uniformly at random sample a full weight vector $\beta \in (\F_3^\times)^m$, suppose we sample $\beta = [1, 1, \cdots , 1, 1]$. We then have $v_1 = \beta$ and $v_2 = \beta + e_i^m = [1,\cdots,1,2,1,\cdots,1]$.
We then sample the required public and secret information $\mathcal{P}$ and $\mathcal{S}$ and generate and send the queries
$$Q_1 = D_1 + E_1 + v_1 \otimes \Delta_1,\ 
Q_2 = D_2 + E_2 + v_2 \otimes \Delta_2.$$
The server responds with $A^i = [A_1 | A_2] = [X\cdot Q_1 | X \cdot Q_2]$.
Individually decoding the response, the user is able to retrieve $$R_1 = \sum_{k=1}^m X^k \text{ and } R_2 = \sum_{k=1}^m X^k + X^i$$ and therefore ultimately $X^i = R_2 - R_1$, the desired file.

The achieved rate as approximated in Corollary \ref{approxRate} is $$R \approx \frac{\delta}{2ns}.$$
\end{example}

\subsection{Security}
\label{SecurityofCBC}

\paragraph{Information set decoding:}
As in the HHW scheme, information set decoding provides an obvious way to attack this scheme. The work factor $\text{Wf} = k^3\binom{n}{k}$ grows super-polynomially in the input parameters $n,k$ of the chosen random linear code. Therefore, for a suitable parameter choice, the scheme is $\epsilon$-secure against a polynomially bounded adversary.

\paragraph{CB-cPIR scheme vs. subquery attack:}
Let us now see in more detail how this scheme circumvents the subquery attack in \cite{bordage2020privacy}.

For the original queries in the HHW scheme\cite{holzbaur2020isit},
$$Q^i = D + E + e^i \otimes \Delta,$$
it was shown in \cite{bordage2020privacy} that for a desired file $X^i$, the submatrices of the query have $\F_q$-rank $$\rank(Q^i[j]) = \rank(D[j]+E[j]) + \delta$$ for $j \neq i$ and $$\rank(Q^i[i]) = \rank(D[i]+E[i]) \leq ns - \delta.$$
These submatrices with high probability have a discernible rank difference, allowing the server to reveal the desired file index.

Consider the case of the CB-cPIR scheme with queries  
$$Q_j = D_j + E_j + v_j \otimes \Delta_j.$$
The submatrices of the query have $\F_q$-rank $$\rank(Q_j[k]) = \rank(D_j[k]+E_j[k]) + \delta$$ for all $k\in [m]$. Therefore, the server cannot ascertain the desired file index by computing the submatrix ranks.

\begin{remark} Since the public and secret information are chosen independently and randomly for each query, the server \textbf{cannot} reconstitute the queries as $Q_1 + Q_2$ to give a HHW query in order to then successfully perform the submatrix rank attack.
\end{remark}

\paragraph{CB-cPIR scheme vs. modified subquery attack:}
\label{modattack}
A natural way to extend the attack in \cite{bordage2020privacy} to the CB-cPIR scheme could be to compute the $\F_q$-ranks of submatrices $Q_j[J]$, where $J \subset [m]$ and $|J| = \wt(v_j)$. For all such $J$ we have
$$\rank(Q_j[J]) \leq (m-\wt(v_j))\delta. $$
Let $\mathcal{I} = \supp(v_j)$. Then
$$\rank(Q_j[\mathcal{I}]) = \rank(D_j[\mathcal{I}]+E_j[\mathcal{I}]) \leq ns - \delta.$$
Otherwise, for $J \neq \supp(v_j)$,
$$\rank(Q_j[J]) = \rank(D_j[J]+E_j[J]) + \delta\leq ns.$$

The support of $v_j$ is only discernible by the attacker if $\rank(D_j[J] + E_j[J])$ does not shrink too much with respect to that of $\mathcal{I}$. 
If we construct $v_j$ such that $(m-\wt(v_j))\delta < ns - \delta$ then $\rank(Q_j[\mathcal{I}])$ and $\rank(Q_j[J])$ are indistinguishable. 
That is, we want $v_j$ such that $$\wt(v_j) \geq m + 1 - \frac{1}{2 R_{\pir}}.$$
\begin{remark}
    We can always sample $\beta$ in a way such that $v_j$ satisfies the above inequality. Effectively, we can sample $\beta$ such that $\wt(v_j) = m$. 
\end{remark}

\begin{proposition}
\label{ProbOfFailure}
    Suppose $Q = D + E + v\otimes \Delta$ is a part of a CB-cPIR query constructed using a vector $v \in \F_q^m$, where $\wt(v) < m + 1 - \frac{1}{2 R_{\pir}}$. Then, for a set $J \subseteq [m] \setminus \supp(v)$ and $|J| = \wt(v)$ we have, 
    $$\prob(\rank(Q[J]) \leq ns - \delta)=\prob(\rank(D[J] + E[J]) \leq ns - 2\delta) \leq \binom{ns - \delta}{ns-2\delta}_q q^{-\delta^2(m-\wt(v))}.$$
    \begin{proof}
        Notice that the rows of $D+E$ are vectors chosen uniformly at random from  $\mathcal{U} = C ~ \oplus ~ \phi_{\bar I}(V^{n-k})$. Keeping notation consistent with \cite{bordage2020privacy}, we represent the set of rows of $D[J] + E[J]$ (seen as vectors of length $ns$ over $\F_q$) by $\rows(D[J] + E[J])$.

        The probability we want to compute is hence

        $$p := \prob(\exists \mathcal{A} \subset \mathcal U, \dim(\mathcal{A}) = ns - 2\delta ~|~  
        \forall y\in \rows(D[J]+E[J]), y \in \mathcal{A}).$$

        By the union bound, we have
        \begin{align*}
            p & \leq \sum_{\mathcal{A}\in Gr_{\mathcal{U}}(ns-2\delta)} \prob(\forall y\in \rows(D[J]+E[J]), y \in \mathcal{A}) \\
            & \leq \sum_{\mathcal{A}\in Gr_{\mathcal{U}}(ns-2\delta)} \prod_{t=1}^{(m - \wt(v))\delta}\prob(y \in \mathcal{A} | y \leftarrow \mathcal{U}) \\
            & \leq \binom{ns - \delta}{ns-2\delta}_q q^{-\delta^2(m-\wt(v))},
        \end{align*}
        where $Gr_{\mathcal{U}}(ns-2\delta)$ denotes the set of $(ns-2\delta)$-dimensional subspaces included in $\mathcal{U}$.
    \end{proof}
\end{proposition}

A rough upper bound for the Gaussian binomial coefficient $\binom{ns - \delta}{ns-2\delta}_q$ is $q^{(\delta + 1)(ns - 2\delta)}$, giving us 
$$p < q^{(\delta + 1)(ns - 2\delta) - \delta^2(m-\wt(v))}.$$
        
This upper bound is meaningful when $(\delta + 1)(ns - 2\delta) \leq \delta^2(m-\wt(v))$. That is, an attacker can distinguish between $Q[\supp(v)]$ and $Q[J]$, where $J\neq \supp(v), |J|=\wt(v)$, with high probability when
$$ m-\wt(v) \geq \left(\frac{\delta+1}{\delta}\right)\left(\frac{1}{2R_{\pir}} - 2\right).$$

\begin{lemma}
\label{BetaAlgo}
    Let $Q = D + E + \beta \otimes \Delta$ be a part of a CB-cPIR query constructed using a vector $\beta \in (\F_q^\times)^m$. Then there exists an algorithm running in $\mathcal{O}((q-1)^h)$ operations over $\F_q$, where $h \geq \left(\frac{\delta+1}{\delta}\right)\left(\frac{1}{2R_{\pir}} - 2\right)$, which can determine the vector $\beta$ with probability 
    $$p > (1 - q^{(\delta + 1)(ns - 2\delta) - \delta^2h})^{\ceil{\frac{m}{h}}}$$
    \begin{proof}
        Let $\mathcal{P}$ be a collection of subsets of cardinality $h$ that cover $[m]$, $|\mathcal{P}| = \ceil{\frac{m}{h}}$. The algorithm consists of the following:
        For each subset $H = \lbrace H_1, \dots, H_h\rbrace \in \mathcal{P}$ return a vector (if unique) $\hat{b} = \phi_{H}(b) \in \F_q^m$ where $b = (b_1,\dots,b_h) \in (\F_q^\times)^h$ such that $$\rank(Q - \hat{b} \otimes b_1^{-1}Q[[m]\setminus \lbrace H_1\rbrace]) \leq ns - \delta.$$

        Notice that the matrix $\mathcal{Q} = Q - \hat{b} \otimes b_1^{-1}Q[[m]\setminus \lbrace H_1\rbrace]$ is of the form 
        $$\mathcal{Q} = D' + E' + (\beta-\hat{b}) \otimes \Delta,$$
        where the rows of $D' + E'$ are vectors from  $\mathcal{U} = C ~ \oplus ~ \phi_{\bar I}(V^{n-k})$. Indeed, we have $\rank(\mathcal{Q}) \leq ns-\delta$ if $\supp(\beta - \hat{b}) = [m]\setminus H$. 
        
        Otherwise, for $\supp(\beta - \hat{b}) \neq [m]\setminus H$: by proposition \ref{ProbOfFailure}. we have $\rank(\mathcal{Q}) \leq ns-\delta$ with negligible probability $$p < q^{(\delta + 1)(ns - 2\delta) - \delta^2h}.$$ 
        Therefore, for any $H \in \mathcal{P}$ the algorithm can determine the $h$ coordinates of $\beta$ indexed by $H$ with probability $p > 1 - q^{(\delta + 1)(ns - 2\delta) - \delta^2h}$. 

        Jointly for all $H \in \mathcal{P}$, the algorithm can determine $\beta$ with probability $$p > (1 - q^{(\delta + 1)(ns - 2\delta) - \delta^2h})^{\ceil{\frac{m}{h}}}.$$ 

        The algorithm involves computing the $\F_q$-rank of $\ceil{\frac{m}{h}}(q-1)^h$ matrices generated by the choice of $\hat{b} \in \F_q^m$ with support $H$ for each $H \in \mathcal{P}$, which amounts to $(q-1)^h \ceil{\frac{m}{h}}h m(ns)^3$ operations over $\F_q$, the algorithm therefore runs in  $\mathcal{O}((q-1)^h)$ operations over $\F_q$.
    \end{proof}
\end{lemma}

\begin{theorem} \label{attackcomplexitymodified}
    Let $\mathcal{Q}^i = [Q_1 ~|~Q_2]$ be a CB-cPIR query. Then there exists an algorithm running in  
    $\mathcal{O}((q-1)^h)$ operations over $\F_q$ which can discern the desired file index $i$ when given as input $\mathcal{Q}^i$ with probability 
    $$p > (1 - q^{(\delta + 1)(ns - 2\delta) - \delta^2(m-1)})(1 - q^{(\delta + 1)(ns - 2\delta) - \delta^2h})^{\ceil{\frac{m}{h}}}.$$
    
    \begin{proof}
        The algorithm first determines the vector $\beta \in (\F_q^\times)^m$ from $Q_1$ by use of the algorithm in lemma \ref{BetaAlgo}. It can then compute $\mathcal{Q} = Q_2 - \beta \otimes \beta_1^{-1}Q_2[[m]\setminus \{1\}]$. The matrix $\mathcal{Q}$ is of the form $\mathcal{Q} = D_2' + E_2' + (e^m_i) \otimes \Delta_2$. The original subquery attack \cite{bordage2020privacy} can then be performed on this matrix in $\mathcal{O}(m^2(ns)^3)$ operations in $\F_q$ with success probability $p > (1 - q^{(\delta + 1)(ns - 2\delta) - \delta^2(m-1)})$. 
        The probability that the algorithm is successful in determining $i$ is the joint probability of success of the two algorithms employed. The number of operations is dominated by the former algorithm.
    \end{proof}
\end{theorem}

\begin{corollary}
    A CB-cPIR query is $(T,\epsilon)$-secure against an adversary running in time $T < \mathcal{O}((q-1)^h)$.
\end{corollary}

\begin{remark}
    Notice that the running time of the above attack is exponential in $h$, which satisfies the inequality $h \geq \left(\frac{\delta+1}{\delta}\right)\left(\frac{1}{2R_{\pir}} - 2\right)$. To achieve adequate security we can always increase the lower bound on $h$ at the cost of a reducing the PIR rate of the scheme. 
\end{remark}

\subsection{Parameter choices}
\label{parameters}

We instantiate CB-cPIR with carefully chosen parameters that maximize the PIR rate while ensuring adequate security against adversaries employing a variety of attacks. 

To counteract the attack described in Section \ref{modattack}, parameters $n,k,s,\text{ and }v$ must be chosen to ensure a sufficiently large lower bound on $h$, enhancing security. However, increasing this lower bound leads to a reduction in the PIR rate of the scheme. Alternatively, to achieve a higher rate, the field size $q$ can be enlarged, increasing the number of possible values for $\beta$. 

Our choice of the dimension $s$ of the extension field must be sufficiently large to prevent an attacker from successfully guessing the subspace $V$ or any subspace containing $V$. From \cite[Lem.~1] {holzbaur2020isit}, we see that the number of guesses required to guess such a subspace is $\binom{s}{s-1}_q \cdot \binom{s-v}{s-v-1}_q^{-1}$. 

The parameters $[n,k]$ of the random code are chosen to ensure security against an attacker performing an information set decoding attack.

Increasing the code length $n$, the field size $q$ or the dimension $s$ of the extension field introduces higher computational complexity. This is due to the increased cost of arithmetic operations over the extension field, which the server must perform to generate responses. While larger fields can enhance security and PIR rate, this trade-off necessitates careful parameter selection to balance performance and computational overhead in practical implementations.

Table \ref{table1} presents carefully selected parameters chosen to ensure privacy, optimize the PIR rate, and minimize computational costs. These parameters reflect a balance between security and efficiency.

\begin{table}[H]
\resizebox{\textwidth}{!}{
\begin{tabular}{|cccccc|c|ccc|}
\hline
\multicolumn{6}{|c|}{Parameters} & Rate (Cor. \ref{approxRate}) & \multicolumn{3}{c|}{Security level (in bits)} \\ \cline{1-10}
$q$ & $s$ & $v$ & $n$ & $k$ & $\delta$ & $R_{\pir}$ & \multicolumn{1}{c|}{ISD Attack} & \multicolumn{1}{c|}{Section \ref{modattack}} & Subspace Attack \\ \hline
$32$ & $32$ & $31$ & $100$ & $50$ & $50$ & $1/128$ & \multicolumn{1}{c|}{$113$} & \multicolumn{1}{c|}{$312$} & $155$ \\
$32$ & $32$ & $30$ & $100$ & $50$ & $100$ & $1/64$ & \multicolumn{1}{c|}{$113$} & \multicolumn{1}{c|}{$153$} & $150$ \\
$2^{16}$ & $12$ & $10$ & $100$ & $50$ & $100$ & $1/24$ & \multicolumn{1}{c|}{$113$} & \multicolumn{1}{c|}{$175$} & $160$ \\
$2^{32} - 5$ & $6$ & $4$ & $120$ & $60$ & $120$ & $1/12$ & \multicolumn{1}{c|}{$133$} & \multicolumn{1}{c|}{$128$} & $128$ \\
$2^{32}$ & $5$ & $3$ & $100$ & $50$ & $100$ & $1/10$ & \multicolumn{1}{c|}{$113$} & \multicolumn{1}{c|}{$128$} & $96$ \\
$2^{61} - 1$ & $6$ & $2$ & $100$ & $50$ & $200$ & $1/6$ & \multicolumn{1}{c|}{$113$} & \multicolumn{1}{c|}{$128$} & $122$ \\ \hline
\end{tabular}
}
\caption{Parameter choices for CB-cPIR}
\label{table1}
\end{table}

\begin{remark} \label{newattack}
    The CB-cPIR scheme as outlined in this paper has been broken by an attack \cite{lage2025securitycodebasedpirscheme} which exploits the repeated use of the same matrix $\Delta$ and by observing rank differences in a specific auxiliary sub-matrix of the query when appended with arbitrary linear combinations of particular rows of the query matrix. This allows the attacker to completely determine the vector $\beta$ in time polynomial in the security parameters, and consequently determine the desired file index $i$ hidden in $\beta + e^m_i$. This attack depends on independently guessing symbols of $\beta$ and therefore the complexity scales linearly in $q$. For all relevant choices of $q$ as specified in Table \ref{table1} the security is severely reduced. 
    \\\\
    This attack can easily be circumvented by replacing $$\beta \in (\F_q^\times)^m \text{ with }S = \begin{bmatrix}
        S_1 \\
        \vdots \\
        S_m
    \end{bmatrix},$$ where $S_j$'s are chosen uniformly at random from $\F_q^{\delta \times \delta} \setminus\{\boldsymbol{0}_{\delta \times \delta}\}$. 
    \\\\
    The Kronecker product $\beta \otimes \Delta_1$ is replaced by $$(I_{m\times m} \otimes \Delta_1) S = \begin{bmatrix}
        S_1 \Delta_1\\
        \vdots \\
        S_m \Delta_1
    \end{bmatrix},$$
    and the Kronecker product $(\beta + e^m_i)\otimes \Delta_2$ is replaced by 
    $$(I_{m\times m} \otimes \Delta_2) (S + e^m_i \otimes I_{\delta\times \delta}) = \begin{bmatrix}
        S_1 \Delta_2\\
        \vdots \\
        (S_i + I_{\delta \times \delta})\Delta_2 \\
        \vdots \\
        S_m \Delta_2
        \end{bmatrix}.$$
    The retrieval happens as usual, from the first query we are able to retrieve $\sum X_jS_j$ and from the second query we are able to retrieve $\sum X_j S_j + X_i$. Subtracting the two gives the desired file. The query and response sizes are unaltered by this modification to the scheme and therefore we maintain competitive rates.
    The server side costs and decoding cost for the user also remain the same as before. The only increase in complexity comes during the query generation phase, where instead of $\beta \otimes \Delta_1$ the user must compute $(I_{m\times m} \otimes \Delta_1) S$. This added complexity in query generation is not overwhelmingly large and can also majorly be done beforehand, independent of knowing which file is desired.
    
    This modification circumvents the attack in \cite{lage2025securitycodebasedpirscheme} by making sure that $\Delta$ is scrambled in each block of the query matrix. Thereby preventing the construction of an auxiliary matrix which allows for predictable rank differences. 

    Further, guessing symbols of $\beta$ is replaced by having to guessing the $S_j$ matrices. This amounts to an astronomical number of guesses $q^{\delta^2}$ for our chosen parameters. This not only makes a modification of the attack in \cite{lage2025securitycodebasedpirscheme} infeasible, but also increases the attack complexity of the modified attack in Theorem \ref{attackcomplexitymodified} from $\mathcal{O}((q-1)^h)$ to roughly $\mathcal{O}((q^{\delta^2}-1)^h)$. This would in fact allow us to maintain small field sizes without largely sacrificing the rate of the scheme.
\end{remark}

\subsection{Extensions}
\label{sec:extensions}
The PIR scheme we presented is suitable for deployment in cases where the size of the files is much larger than the number of files stored on the database. In many practical scenarios this may not be the case. We therefore extend our construction to handle files of smaller size to make it applicable in other realistic deployment scenarios.

\subsubsection{Square database}
The first instance of a single server PIR scheme \cite{kushilevitz1997replication} to have nontrivial communication made use of the ``square database'' approach. Here a database consisting of $m$ files is reshaped into a $\sqrt{m} \times \sqrt{m}$ square matrix of files and stored on the server. The user, who desires the $i^{th}$ file in the database decomposes the index $i \in [m]$ into the pair of coordinates $(i_{\text{row}}, i_{\text{col}}) \in [\sqrt{m}] \times [\sqrt{m}]$. The user then builds a query to privately retrieve column $i_{\text{col}}$ of the square file matrix. From the retrieved column the user can then isolate the row $i_{\text{row}}$ to obtain their desired file. We can use this simple notion, and use CB-cPIR on a reshaped, square database to improve rates for files of small size.

In this form of deployment, the user desires, as before a file of size $L\delta \log_2(q)$. The size of the query uploaded by the user will be $2\sqrt{m}\delta n \log_2(q^s)$. And the size of the server answer will be $2L\sqrt{m} n \log_2(q^s)$. This, as per corollary \ref{approxRate1} gives us the rate $$R_{\pir} = \frac{L\delta}{2ns\sqrt{m}(\delta + L)}.$$
The unextended scheme has communication linear in the number of files in the database, which when used for retrieval of small files results in a PIR scheme, which is less efficient than trivially downloading the entire database. In contrast, this version of the scheme has communication sublinear in the number of files on the database, allowing for a better than trivial efficiency even in the case of small files. For example, to retrieve a file of maximum size $log_2(q)$ bits the rate of the PIR scheme is $R_{\pir} = \frac{1}{4ns\sqrt{m}}$.

In the following section the idea of decreasing the size of the query is extended by viewing the database as a $t$-dimensional hypercube. 

\subsubsection{Iterative use}
When the size of the files is small with respect to the number of files in the database the upload cost, \emph{i.e.}, the size of the query may dominate the communication cost. We extend the CB-cPIR scheme to retrieve the desired file after $t$ iterations from a reshaped database, allowing us to reduce the total upload cost of the scheme.

\paragraph{Database:}
The database remains the single server, which is represented by $X = [X^0 \cdots X^{m-1}] \in \F_q^{L\times m\delta}$. Suppose $m = x^t$ for some integers $x$ and $t$. 

\begin{definition}
    Define a re-indexing function $\mathcal{F}_r: [0:m-1] \to [0:x-1]^{t-r+1}$, which maps $$i \mapsto \left(\floor*{\frac{i}{x^{r-1}}} \mod x, \dots , \floor*{\frac{i}{x^{t-1}}} \mod x \right).$$
\end{definition}

Note that the image of this map has a natural ordering inherited from the natural ordering of $[0:m-1]$.
\\We now work with the re-indexed database represented by $X' = [X^{\mathcal{F}_1(0)} \cdots X^{\mathcal{F}_1(m-1)}]$, which can be imagined as a t-dimensional cube of files.

\begin{definition}
    Define some bijective function $$\boldsymbol{\delta}_M: \F_q^{M \times 2ns} \to \F_q^{M\frac{2ns}{\delta} \times \delta}.$$
\end{definition}

Now, suppose the user wants to retrieve the $d^{th}$ file ($\mathcal{F}_1(d) = (d_1, \dots , d_t)$).

\paragraph{Iterations:}

Initially, we have $X'_1 = [X_1^{\mathcal{F}_1(0)} \cdots X_1^{\mathcal{F}_1(m-1)}]$.
\\\\
\textbf{Round} $r$:
Define $X_r = [X_r^0 \cdots X_r^{x-1}] \in   \F_q^{L(\frac{2ns}{\delta})^{r-1}x^{t-r} \times x\delta}$ where $X_r^i$ is the naturally ordered stack of all files $X^I \in X'_r$ such that $I[1]=i$.
The user then constructs a query $Q^{d_r}$ as prescribed by our scheme to retrieve $X_r^{d_r}$. The server then computes the answer $X_r \cdot Q^{d_r}$.
\\We can also view $X_r$ as $$X_r = \begin{bmatrix}
    B_r^{\mathcal{F}_r(0)} \\
    B_r^{\mathcal{F}_r(x^r)} \\
    \vdots \\
    B_r^{\mathcal{F}_r((x^{t-r}-1)x^r)}
\end{bmatrix}.$$
Where $B_r^I \in \F_q^{L(\frac{2ns}{\delta})^{r-1} \times x\delta}$ is the naturally ordered vector of all files $X_r^{\mathcal{F}_r(j)} \in X'_r$ such that $\mathcal{F}_{r+1}(j) = I$.
\\The server response is then 
$$ X_r\cdot Q^{d_r} = \begin{bmatrix}
    B_r^{\mathcal{F}_r(0)} \\
    B_r^{\mathcal{F}_r(x^r)} \\
    \vdots \\
    B_r^{\mathcal{F}_r((x^{t-r}-1)x^r)}
\end{bmatrix} \cdot Q^{d_r} = \begin{bmatrix}
    B_r^{\mathcal{F}_r(0)} \cdot Q^{d_r}\\
    B_r^{\mathcal{F}_r(x^r)} \cdot Q^{d_r}\\
    \vdots \\
    B_r^{\mathcal{F}_r((x^{t-r}-1)x^r)} \cdot Q^{d_r}
\end{bmatrix}
.$$

At the end of each round, we store statefully on the server 
\begin{multline*}
    X'_{r+1} = [\boldsymbol{\delta}_{L(\frac{2ns}{\delta})^{r-1}}(B_r^{\mathcal{F}_r(0)} \cdot Q^{d_r}) \cdots \\
    \cdots \boldsymbol{\delta}_{L(\frac{2ns}{\delta})^{r-1}}(B_r^{\mathcal{F}_r((x^{t-r}-1)x^r)} \cdot Q^{d_r})] \\
    = [X_{r+1}^{\mathcal{F}_r(0)} \cdots X_{r+1}^{\mathcal{F}_r((x^{t-r}-1)x^r)}] 
    \in \F_q^{L(\frac{2ns}{\delta})^{r} \times x^{t-r}\delta}.
\end{multline*}

After completing $\omega$ rounds on the $t$-dimensional database, the user downloads the final $R_1 = X'_{\omega+1} \in \F_q^{L(\frac{2ns}{\delta})^{r} \times x^{t-\omega}\delta}$.

\paragraph{Decoding the response:}
We decode the response over $\omega$ rounds, reshaping each round appropriately to $M \times ns$ before using the \textbf{Recover} function as specified in the CB-cPIR construction in Fig. \ref{CBCpirAlgo}.
\\\textbf{Round} $r$:
\begin{align*}
    R_{r+1} &=\textbf{ Recover}(\boldsymbol{\delta}^{-1}(R_r))\\
    &= X_{t-r+1}^{(d_{t-r+1}, \dots d_t)}= \boldsymbol{\delta}([X_{t-r}^{(0,d_{t-r+1}, \dots d_t)} \cdots X_{t-r}^{(x-1,d_{t-r+1}, \dots d_t)}]\cdot Q_{d-r})
\end{align*}
Here, $Q_j$ represents the query sent during the $j^{th}$ round, and $Q_0$ is just the identity matrix.
After $\omega$ rounds we finally have $R_{\omega+1} = X_{t-\omega+1}^{(d_{t-\omega+1}, \dots d_t)} = X_{t-\omega+1}^{\mathcal{F}_\omega(d)}$, which consists of all files $X^I$ such that the last $\omega$ coordinates of $I$ agree with $\mathcal{F}_{t-\omega}(d)$. 
The user can then extract the file $X^{\mathcal{F}_1(d)}$, which is the desired file, from the recovered response.

\paragraph{Rate of the iterative scheme:}
To retrieve a file of size $L\delta \log_2(q)$ bits we determine the communication costs for the iterative form of deployment of CB-cPIR.

Concretely, for a database viewed as a $t$-dimensional hypercube, the size (in bits) of each of the $\omega$ uploaded queries is $2x\delta ns \log_2(q)$, which amounts to a total upload cost of 
$$C_{up} = 2x\omega \delta ns = 2m^{\frac{1}{t}}\omega \delta ns \log_2(q).$$

The total download cost is given by the size of the final response $R_1$,
$$C_{down} = L(\frac{2ns}{\delta})^{\omega} x^{t-\omega}\delta \log_2(q) = L(\frac{2ns}{\delta})^{\omega} m^{\frac{t-\omega}{t}}\delta \log_2(q).$$

This gives us the PIR rate of the scheme: 
$$R_{\pir} = \frac{L\delta}{ 2m^{\frac{1}{t}}\omega \delta ns + L(\frac{2ns}{\delta})^{\omega} m^{\frac{t-\omega}{t}}\delta}.$$

\paragraph{Computational complexity of the iterative scheme:}
We now concretely determine the total computational costs incurred by the server and user in the private retrieval of a single file when using the iterative version of CB-cPIR. 

\paragraph{Server complexity:}
Cumulatively over $\omega$ rounds, the concrete cost of multiplying the query matrices with the statefully stored databases, is $2Lm\delta ns \left( \frac{(2ns/\delta m^{\frac{1}{t}})^\omega -1}{(2ns/\delta m^{\frac{1}{t}}) -1} \right)$ multiplications in $\F_q$.

\paragraph{User complexity:}
The user complexity for the PIR protocol is divided into two parts, the complexity of generating the query and the complexity of decoding the server response.
\begin{itemize}
    \item Query generation: The complexity of query generation is dominated by the following steps in the protocol: \begin{itemize}
        \item Generating the random codeword matrix $D$: This requires $m^{1/t}\delta kns$ multiplications over $\F_q$. 
        \item Generating the $V$ noise matrix $E$: This requires $m^{1/t}\delta kvs$ multiplications over $\F_q$.
        \item Generating the $W$ noise matrix $\Delta$: This requires $\delta kws$ multiplications over $\F_q$.     
        \item Kronecker product: This requires $\min(m^{1/t},q)\delta k s$ multiplications over $\F_q$. The term $\min(m^{1/t},q)$ arises from the fact that when $q > m^{1/t}$ there is a pigeon-holing of scalar multiplications in the Kronecker product. 
    \end{itemize}
    The above computational costs are incurred twice, once for each part of the query matrix for a total of $\omega$ queries. The total complexity for query generation is then:
    $$C_{Qgen} = 2\omega\delta k s(m^{1/t}n + m^{1/t}v + w + \min(m^{1/t},q)).$$
    \item Decoding response: The decoding of the response is dominated by the following steps in the protocol:
    \begin{itemize}
        \item Erasure decoding the response: This involves \\$Lkns\left(\sum_{r=1}^\omega (\frac{ns}{\delta})^{r-1}m^{\frac{t-r}{t}} \right) = Lm^{\frac{t-1}{t}}kns \left( \frac{(ns/\delta m^{\frac{1}{t}})^\omega -1}{(ns/\delta m^{\frac{1}{t}}) -1} \right)$ multiplications over $\F_q$.
        \item Projection onto the subspace $W$: This involves viewing each $\F_{q^s}$ element as a vector representation in terms of our chosen secret basis $\Gamma$. This requires $Lm^{\frac{t-1}{t}}k\left( \frac{(ns/\delta m^{\frac{1}{t}})^\omega -1}{(ns/\delta m^{\frac{1}{t}}) -1} \right)(s^2 + w)$ multiplications over $\F_q$.
        \item Inverting the matrix $\Delta$: This involves $Lm^{\frac{t-1}{t}}k\left( \frac{(ns/\delta m^{\frac{1}{t}})^\omega -1}{(ns/\delta m^{\frac{1}{t}}) -1} \right)\delta s$ multiplications over $\F_q$.
    \end{itemize}
    The above computational costs are incurred twice, once for each part of the response matrix for a total of $\omega$ rounds. The total complexity to decode the response is then:
    $$C_{Adec} = 2\omega Lm^{\frac{t-1}{t}}k\left( \frac{(ns/\delta m^{\frac{1}{t}})^\omega -1}{(ns/\delta m^{\frac{1}{t}}) -1} \right)(ns + s^2 + w + \delta s).$$
\end{itemize}

\begin{remark}
    When the iterative version of the scheme is considered with $t = 2$ and $\omega = 1$, it coincides precisely with the case of considering a square database. 
\end{remark}

\section{Comparisons}
\label{sec:comparison}
In this section, we provide a comprehensive analysis of some well-known computational PIR schemes that leverage the hardness of lattice-based problems in comparison to CB-cPIR, which leverages a hard problem in coding theory. Our evaluation focuses on three critical aspects: communication costs, which quantify the amount of data exchanged between the client and server during query execution; computational complexity, which measures the computational effort required by the server to generate responses; and the PIR rate, which reflects the efficiency of data retrieval as a ratio of retrieved data to communication overhead. By systematically comparing these schemes, we aim to highlight their trade-offs and performance characteristics, offering insights into their suitability for various practical applications. 

\begin{remark}
    In our analysis, when determining the computational complexity, for all schemes we consider naïve matrix multiplications without any optimizations. Further, we simply count the number of multiplications over comparable fields/rings for valid comparisons.
\end{remark}

Below, we give appropriate parameter choices for XPIR and SimplePIR, for a desirable level of security.

\begin{table}[H]
    \centering
    \begin{tabular}{|c|c|c|c|c|c|}
        \hline
        \multicolumn{2}{|c|}{Parameters} & Maximum & Plaintext size & Ciphertext size & Expansion factor \\
        \cline{1-2}
        $n$ & $\log(q)$ & Security (in bits) & $s_p$ & $s_c$ &  $s_p/s_c$ \\  
        \hline
        $1024$  & $\approx 60$  & $97$  & $\leq 20$ Kbits  & $128$ Kbits  & $\geq 6.4$ \\
        $2048$  & $\approx 120$ & $91$  & $\leq 100$ Kbits & $512$ Kbits  & $\geq 5.12$ \\
        $4096$  & $\approx 120$ & $335$ & $\leq 192$ Kbits & 1 Mbit     & $\geq 5.3$ \\
        \hline
    \end{tabular}
    \caption{Parameters for XPIR}
    \label{Xparam}
\end{table}

\begin{table}[H]
    \centering
    \begin{tabular}{|c|c|c|c|c|c|}
        \hline
        Database size & \multicolumn{3}{|c|}{Parameters} & Maximum \\
        \cline{2-4}
        (in bits) & n & Modulus $q$ & Plaintext modulus $p$  & Security (in bits) \\  
        \hline
        $2^{26}$ & $1024$ & $2^{32}$ & $991$ & $128$  \\
        $2^{34}$ & $1024$ & $2^{32}$ & $495$ & $128$  \\
        $2^{42}$ & $1024$ & $2^{32}$ & $247$ & $128$  \\
        \hline
    \end{tabular}
    \caption{Parameters for SimplePIR}
    \label{Simpleparam}
\end{table}

\subsection{XPIR}

XPIR \cite{aguilar2016xpir} is a computational Private Information Retrieval protocol that inherits its security from Ring Learning with Errors (RLWE), a post-quantum cryptographic problem. The protocol relies on an additively-homomorphic building block from the fully homomorphic encryption scheme in \cite{FHE-RLWE}, enabling computations on encrypted data without decryption. This ensures that the server cannot infer any information about the user's query, as it processes the encrypted request directly. Each query is represented as a polynomial in the ring $R_q = \frac{\mathbb{Z}_q[X]}{\langle X^n - 1 \rangle}$, where $n$ decides the security level of the scheme. 

In our comparison, we will consider the XPIR protocol instantiated with the parameters $(n, \log(q)) = (1024, 60)$ as given in the first row of Table \ref{Xparam} against the CB-cPIR protocol instantiated with the parameters $(q, n, k, s, v) = (2^{61}-1, 100, 50, 6, 2)$ as given in the last row of Table \ref{table1}. These choices of parameters provide a maximum security of $97$ bits in the XPIR protocol, and a maximum security of $113$ bits in the CB-cPIR protocol.

\subsubsection{Communication cost and rate}
The query sent to the server consists of $m$ ciphertexts each of size $s_c$ as determined by the parameter $n$ of the RLWE homomorphic encryption scheme for a required level of security. This gives us the upload cost of the scheme $C_{up} = ms_c$. 

Suppose the database contains files of size $L$. 
The server upon receiving the query splits each file into $L/s_p$ plaintext messages, and performs the appropriate homomorphic operations with the query. The server then responds with $L/s_p$ ciphertexts each of size $s_c$. This gives us the download cost $C_{down}=L\frac{s_c}{s_p}$. 

The total cost of communication is then $$C_{total} = C_{up} + C_{down} = ms_c + L\frac{s_c}{s_p}.$$

The rate of the scheme is $$R_{\pir} = \frac{L}{ms_c + L\frac{s_c}{s_p}} = \frac{L s_p}{ms_ps_c + Ls_c} .$$

\begin{figure}[H]
  \centering
  \begin{subfigure}{0.49\textwidth}
    \centering
    \resizebox{\textwidth}{!}{\begin{tikzpicture}
  \begin{axis}[
    xlabel={Size of files},
    ylabel={Rate},
    axis lines = left,
    legend pos = south east,
    domain=6400:10000000000,
    xmode=log,
    ymode=log,
    samples=10
    ]

    \def\m{10000}  
    \def\t{1}    
    \def\w{1}    
    \def\n{100}  
    \def\s{3}    
    \def\d{100}  

    \addplot [red, thick, mark=*]{x / (2 * \m^(1/\t) * \w * \n * \s + x * ((2 * \n * \s / \d)^\w * \m^((\t - \w)/\t))) };
    \addlegendentry{CB-cPIR ($t = 1$, $\omega = 1$)}

    \def\t{2}    
    \def\w{1}    
    
    \addplot [green, thick, mark=*]{x / (2 * \m^(1/\t) * \w * \n * \s + x * ((2 * \n * \s / \d)^\w * \m^((\t - \w)/\t))) };
    \addlegendentry{CB-cPIR ($t = 2$, $\omega = 1$)}
    
    \addplot [blue, thick, mark=*]{x / (\m * 128000 + x * 6.4)};
    \addlegendentry{XPIR}

  \end{axis}
\end{tikzpicture}}
    \caption{Size of files (in bits) vs. rate for database containing $10000$ files}
    \label{rateXvCfixedfiles}
  \end{subfigure}
  \hfill
  \begin{subfigure}{0.49\textwidth}
    \centering
    \resizebox{\textwidth}{!}{\begin{tikzpicture}
  \begin{axis}[
    xlabel={Size of files},
    ylabel={Rate},
    axis lines = left,
    legend pos = south east,
    domain=6400:1250000,
    xmode=log,
    ymode=log,
    samples=10
    ]

    \def\t{1}    
    \def\w{1}    
    \def\n{100}  
    \def\s{3}    
    \def\d{100}  

    \addplot [red, thick, mark=*]{x / (2 * (8000000000/x)^(1/\t) * \w * \n * \s * \d + x * ((2 * \n * \s / \d)^\w * (8000000000/x)^((\t - \w)/\t))) };
    \addlegendentry{CB-cPIR ($t = 1$, $\omega = 1$)}

    \def\t{2}    
    \def\w{1}    
    
    \addplot [green, thick, mark=*]{x / (2 * (8000000000/x)^(1/\t) * \w * \n * \s * \d + x * ((2 * \n * \s / \d)^\w * (8000000000/x)^((\t - \w)/\t))) };
    \addlegendentry{CB-cPIR ($t = 2$, $\omega = 1$)}
    
    \addplot [blue, thick, mark=*]{x / ((8000000000/x) * 128000 + x * 6.4)};
    \addlegendentry{XPIR}

  \end{axis}
\end{tikzpicture}}
    \caption{Size of files (in bits) vs. rate for a database of size $1$GB}
    \label{rateXvCfixedDB}
  \end{subfigure}
  \caption{Comparison of file size vs. rate for different database configurations}
\end{figure}
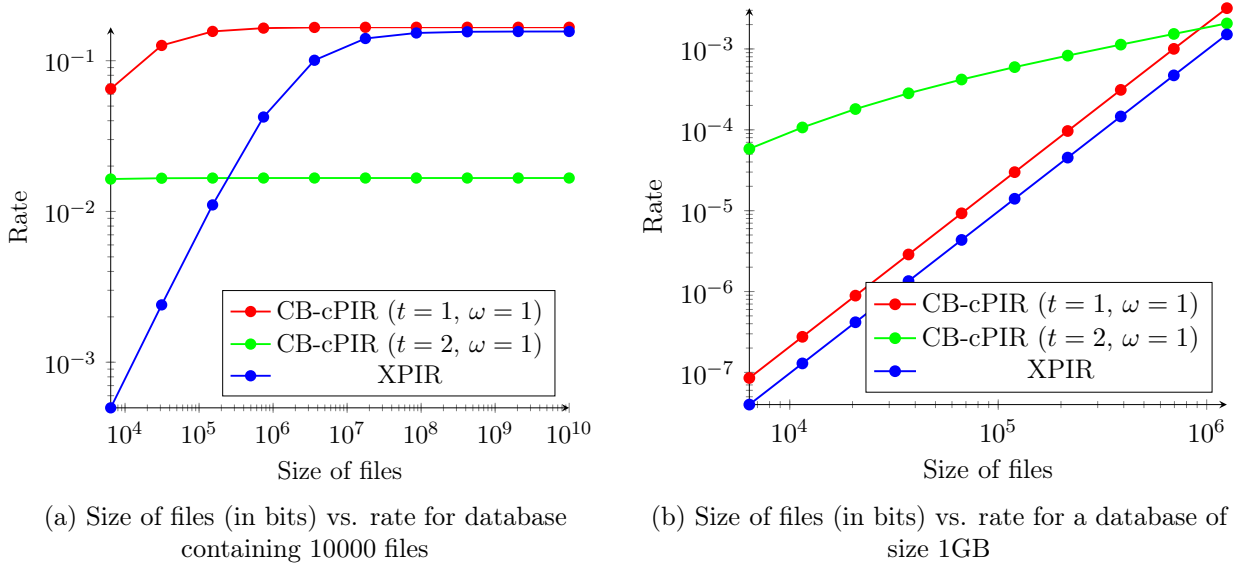

In the figures above, we compare the PIR rate of CB-cPIR with parameters $(q, n, k, s, v) = (2^{61}-1, 100, 50, 6, 2)$ against the PIR rate of XPIR, with parameters $(n, \log(q)) = (1024, 60)$, which give an expansion factor of $s_c/s_p = 6.4$.  

In Fig. \ref{rateXvCfixedfiles}, we fix the number of files and plot the PIR rates as the file sizes increase. The results show that the PIR rate of the schemes asymptotically approaches its optimal value. This occurs because, as file sizes grow, the impact of the upload cost on total communication becomes negligible. In Fig. \ref{rateXvCfixedDB}, we fix the total database size and examine how the PIR rate changes as file sizes increase (or equivalently, as the number of files decreases).  In both scenarios, we see that CB-cPIR has favourable communication overhead in comparison with XPIR.

\subsubsection{Computational complexity}
In this section we concretely determine the computational costs associated with the private retrieval of a file of size $L$ in $\mathbb{Z}_p$ using the XPIR protocol. The computational costs in XPIR are dominated by multiplications of polynomials over the ring $R_q = \frac{\mathbb{Z}_q[X]}{\langle X^n - 1 \rangle}$, the authors reduce the complexity of these multiplications using Number-Theoretic Transform (NTT) for polynomials \cite{NTT} and using precomputed Newton coefficients for modular integer multiplications. Similar optimizations can be used to improve the computational costs associated with multiplications in CB-cPIR. In our comparison, we do not consider these optimizations and determine computational costs based on naïve multiplications.

\textbf{Server complexity:} The concrete cost of generating a response is $mLn^2$ multiplications in $\mathbb{Z}_q$. This is a result of $m\cdot L$ multiplications of polynomials over the quotient ring $R_q = \frac{\mathbb{Z}_q[X]}{\langle X^n - 1 \rangle}$.

\textbf{User complexity:} The user complexity for the XPIR protocol is divided into two parts, the complexity of generating the query and the complexity of decoding the server response.
\begin{itemize}
    \item Query generation: The complexity of query generation is dominated by the computation of $a \cdot s$, which requires $mn^2$ operations over $\mathbb{Z}_q$.
    \item Decoding response: The decoding of the response is dominated by the computation of $\mathrm{Response}\cdot s$, which requires $Ln^2$ operations over $\mathbb{Z}_q$.
\end{itemize}

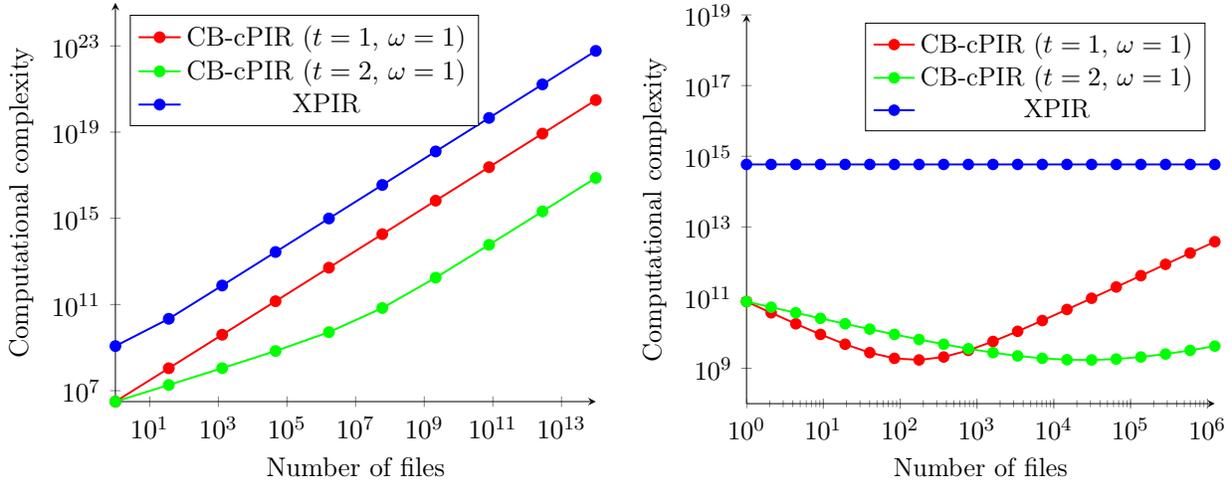
\begin{figure}[H]
  \centering
  \begin{subfigure}{0.49\textwidth}
    \centering
    \resizebox{\textwidth}{!}{\begin{tikzpicture}
  \begin{axis}[
    xlabel={Number of files},
    ylabel={Computational complexity},
    axis lines = left,
    legend pos = north west,
    domain=1:100000000000000,
    xmode=log,
    ymode=log,
    ymax = 10^25,
    samples=10
    ]

    \def\L{1.25}    
    \def\t{1}    
    \def\omegaw{1}    
    \def\w{2}    
    \def\n{100}  
    \def\s{3}    
    \def\d{100}  
    \def\k{50}    
    \def\v{1}    
    \addplot[red, thick, mark=*]{
    2 * \L * x * \n * \s * ( ((2 * \n * \s) / (\d * x^(1/\t)))^\omegaw - 1 ) / ( ((2 * \n * \s) / (\d * x^(1/\t))) - 1 )
    + 2 * \omegaw * ( x^(1/\t) * \d * \k * \n * \s + x^(1/\t) * \d * \k * \v * \s + \d * \k * \w * \s + x^(1/\t) * \d * \k * \s ) 
    + ( 2 * \omegaw * \L / \d ) * x^((\t-1)/\t) * \k * ( ((\n * \s) / (\d * x^(1/\t)))^\omegaw - 1 ) / ( ((\n * \s) / (\d * x^(1/\t))) - 1 ) * ( \n * \s + \s^2 + \w + \d * \s )
    };
    \addlegendentry{CB-cPIR ($t = 1$, $\omega = 1$)}
    
    \def\t{2}    
    \def\omegaw{1}    
    \addplot[green, thick, mark=*]{
    2 * \L * x * \n * \s * ( ((2 * \n * \s) / (\d * x^(1/\t)))^\omegaw - 1 ) / ( ((2 * \n * \s) / (\d * x^(1/\t))) - 1 )
    + 2 * \omegaw * ( x^(1/\t) * \d * \k * \n * \s + x^(1/\t) * \d * \k * \v * \s + \d * \k * \w * \s + x^(1/\t) * \d * \k * \s ) 
    + ( 2 * \omegaw * \L / \d ) * x^((\t-1)/\t) * \k * ( ((\n * \s) / (\d * x^(1/\t)))^\omegaw - 1 ) / ( ((\n * \s) / (\d * x^(1/\t))) - 1 ) * ( \n * \s + \s^2 + \w + \d * \s )
    };
    \addlegendentry{CB-cPIR ($t = 2$, $\omega = 1$)}

    \def\L{560} 
    \addplot[blue, thick, mark=*]{
     (x + \L + x * \L) * 1024 * 1024
    };
    \addlegendentry{XPIR}
    
  \end{axis}
\end{tikzpicture}}
    \caption{Number of files vs. computational complexity for files of size $1$KB}
    \label{compXvCfixedfile}
  \end{subfigure}
  \hfill
  \begin{subfigure}{0.49\textwidth}
    \centering
    \resizebox{\textwidth}{!}{\begin{tikzpicture}
  \begin{axis}[
    xlabel={Number of files},
    ylabel={Computational complexity},
    axis lines = left,
    domain=1:1250000,
    xmode=log,
    ymode=log,
    ymin = 100000000,
    ymax = 10000000000000000000,
    samples=20
    ]

    \def\t{1}    
    \def\omegaw{1}    
    \def\w{2}    
    \def\n{100}  
    \def\s{3}    
    \def\d{100}  
    \def\k{50}    
    \def\v{1}    
    \def\p{20000}    
    \addplot[red, thick, mark=*]{
    2 * (1250000/x) * x * \n * \s * ( ((2 * \n * \s) / (\d * x^(1/\t)))^\omegaw - 1 ) / ( ((2 * \n * \s) / (\d * x^(1/\t))) - 1 )
     + 2 * \omegaw * ( x^(1/\t) * \d * \k * \n * \s + x^(1/\t) * \d * \k * \v * \s + \d * \k * \w * \s + x^(1/\t) * \d * \k * \s ) 
    + ( 2 * \omegaw * (1250000/x)) * x^((\t-1)/\t) * \k * ( ((\n * \s) / (\d * x^(1/\t)))^\omegaw - 1 ) / ( ((\n * \s) / (\d * x^(1/\t))) - 1 ) * ( \n * \s + \s^2 + \w + \d * \s )
    };
    \addlegendentry{CB-cPIR ($t = 1$, $\omega = 1$)}
    
    \def\t{2}    
    \def\omegaw{1}    
    \addplot[green, thick, mark=*]{
    2 * (1250000/x) * x * \n * \s * ( ((2 * \n * \s) / (\d * x^(1/\t)))^\omegaw - 1 ) / ( ((2 * \n * \s) / (\d * x^(1/\t))) - 1 )
    + 2 * \omegaw * ( x^(1/\t) * \d * \k * \n * \s + x^(1/\t) * \d * \k * \v * \s + \d * \k * \w * \s + x^(1/\t) * \d * \k * \s ) 
    + ( 2 * \omegaw * (1250000/x)) * x^((\t-1)/\t) * \k * ( ((\n * \s) / (\d * x^(1/\t)))^\omegaw - 1 ) / ( ((\n * \s) / (\d * x^(1/\t))) - 1 ) * ( \n * \s + \s^2 + \w + \d * \s )
    };
    \addlegendentry{CB-cPIR ($t = 2$, $\omega = 1$)}
    
    \addplot[blue, thick, mark=*]{
     (x + (8000000000/(x*log2(\p)) + x * (8000000000/(x*log2(\p))) * 1024 * 1024
    };
    \addlegendentry{XPIR}

  \end{axis}
\end{tikzpicture}}
    \caption{Number of files vs. computational complexity for a database of size $1$GB}
    \label{compXvCfixedDB}
  \end{subfigure}
  \caption{Comparison of computational complexity for different file sizes and database configurations}
\end{figure}

In the figures above, we compare the computational complexity of CB-cPIR with parameters $(q, n, k, s, v) = (2^{61}-1, 100, 50, 6, 2)$ against the PIR rate of XPIR, with parameters $(n, \log(q)) = (1024, 60)$, which give an expansion factor of $s_c/s_p = 6.4$. Importantly, the modulus $q$ in both protocols is comparable in size ($\approx 61$ bits), which makes our comparisons valid.

In Fig. \ref{compXvCfixedfile}, we fix the size of the files and examine how the computational complexity changes as the number of files increases. In Fig. \ref{compXvCfixedDB} we fix the size of the database and examine how the computational complexity changes as the number of files increases. In both scenarios, CB-cPIR is seen to outperform XPIR in terms of computational costs.

\subsection{SimplePIR}
SimplePIR \cite{SimplePIR} is a computational PIR protocol that inherits its security from the learning with errors (LWE) problem. More specifically, the protocol utilizes the secret key version of Regev's LWE encryption scheme. Regev's encryption using a lattice dimension $n$ involves using an LWE matrix $A \in \mathbb{Z}_q^{m\times n}$, a secret value $s \in \mathbb{Z}_q^n$, an error vector $e$ of length $m$ sampled from a specific error distribution, and the message $\mu \in \mathbb{Z}_p^m$. The message is then encrypted as 
$$\mathrm{Enc}(\mu) = (A, As + e + \floor{q/p}\mu).$$
In SimplePIR, the hint consists of a one-time download of the product of the database with the LWE matrix $A$. This hint can be reused polynomially many times allowing for amortization of the scheme. The query with respect to a desired file index $i \in [m]$ consists of the latter part of the encryption of the standard unit vector $\mu = e_i^m$, for which the server response is the matrix product between the query and the database. Note that this protocol uses the ``square'' approach and $m$ is replaced by $\sqrt{m}$ in the case of the reshaped square database. 

The cost of uploading the matrix $A$ is significantly reduced by compression using a pseudorandom key; therefore, in our analysis we ignore this cost.

In our comparison, we will consider the SimplePIR protocol instantiated with the parameters $(q, p, n) = (2^{32}, 495, 1024)$ as given in the second row of Table \ref{Simpleparam} against the CB-cPIR protocol instantiated with the parameters $(q, n, k, s, v) = (2^{32}-5, 120, 60, 6, 4)$ as given in the fourth row of Table \ref{table1}. These choices of parameters provide a maximum security of $128$ bits in both the SimplePIR and the CB-cPIR protocol.

\subsubsection{Communication cost and rate}
To query and retrieve a file of size $L$ in $\mathbb{Z}_p$, the concrete communication costs are:
\begin{itemize}
    \item One-time hint download: $nL\sqrt{m}$ elements of $\mathbb{Z}_q$.
    \item Per query upload cost: $\sqrt{m}$ elements of $\mathbb{Z}_q$.
    \item Server response: $L\sqrt{m}$ elements of $\mathbb{Z}_q$.
\end{itemize}
The total cost of communication amortized over $t$ queries is then $$C_{total} = nL\sqrt{m} + (L+1)t\sqrt{m}.$$
And the amortized PIR rate is 
$$R_{\pir} = \frac{Lt \log(p)}{(nL\sqrt{m} + (L+1)t\sqrt{m})\log(q)}.$$

\begin{figure}[H]
  \centering
  \begin{subfigure}{0.49\textwidth}
    \centering
    \resizebox{\textwidth}{!}{\begin{tikzpicture}
  \begin{axis}[
    xlabel={Number of files},
    ylabel={Rate},
    axis lines = left,
    legend pos = south west,
    domain=1:1000000000000,
    xmode=log,
    ymode=log,
    samples=20
    ]

    \def\t{1}    
    \def\w{1}    
    \def\n{120}  
    \def\s{6}    
    \def\d{120}  
    \def\tt{100}    
    \def\nn{1024}  
    \def\L{26667}  
    \def\p{247}  
    \def\q{4294967296}  

    \addplot [red, thick, mark=*]{\L / (2 * (x)^(1/\t) * \w * \n * \s + \L* ((2 * \n * \s / \d)^\w * (x)^((\t - \w)/\t))) };
    \addlegendentry{CB-cPIR ($t = 1$, $\omega = 1$)}

    \def\t{2}    
    \def\w{1}    
    
    \addplot [green, thick, mark=*]{\L / (2 * (x)^(1/\t) * \w * \n * \s + \L* ((2 * \n * \s / \d)^\w * (x)^((\t - \w)/\t))) };
    \addlegendentry{CB-cPIR ($t = 2$, $\omega = 1$)}

    \def\t{3}    
    \def\w{2}    
    
    \addplot [yellow, thick, mark=*]{\L / (2 * (x)^(1/\t) * \w * \n * \s + \L* ((2 * \n * \s / \d)^\w * (x)^((\t - \w)/\t))) };
    \addlegendentry{CB-cPIR ($t = 3$, $\omega = 2$)}

    \addplot [blue, thick, mark=*]{(\L * \tt) / ((\nn * \L * sqrt(x) + (\L + 1) * \tt * sqrt(x)) * log2(\q))};
    \addlegendentry{SimplePIR (amortized)}

    \def\tt{1}
    
    \addplot [purple, thick, mark=*]{(\L * \tt) / ((\nn * \L * sqrt(x) + (\L + 1) * \tt * sqrt(x)) * log2(\q))};
    \addlegendentry{SimplePIR (not amortized)}
    
  \end{axis}
\end{tikzpicture}}
    \caption{Number of files vs. rate for database containing files of size $4$KB}
  \end{subfigure}
  \hfill
  \begin{subfigure}{0.49\textwidth}
    \centering
    \resizebox{\textwidth}{!}{\begin{tikzpicture}
  \begin{axis}[
    xlabel={Size of files},
    ylabel={Rate},
    axis lines = left,
    legend pos = south east,
    domain=3200:25000000,
    xmode=log,
    ymode=log,
    samples=10
    ]

    \def\t{1}    
    \def\w{1}    
    \def\n{120}  
    \def\s{6}    
    \def\d{120}  
    \def\tt{100}    
    \def\nn{1024}  
    \def\p{495}  
    \def\q{4294967296}  

    \addplot [red, thick, mark=*]{x / (2 * (8000000000/(x*32))^(1/\t) * \w * \n * \s + x * ((2 * \n * \s / \d)^\w * (8000000000/(x*32))^((\t - \w)/\t))) };
    \addlegendentry{CB-cPIR ($t = 1$, $\omega = 1$)}

    \def\t{2}    
    \def\w{1}    
    
    \addplot [green, thick, mark=*]{x / (2 * (8000000000/(x*32))^(1/\t) * \w * \n * \s + x * ((2 * \n * \s / \d)^\w * (8000000000/(x*32))^((\t - \w)/\t))) };
    \addlegendentry{CB-cPIR ($t = 2$, $\omega = 1$)}

    \def\t{2}    
    \def\w{2}    
    
    \addplot [yellow, thick, mark=*]{x / (2 * (8000000000/(x*32))^(1/\t) * \w * \n * \s + x * ((2 * \n * \s / \d)^\w * (8000000000/(x*32))^((\t - \w)/\t))) };
    \addlegendentry{CB-cPIR ($t = 2$, $\omega = 2$)}
    
    \addplot [blue, thick, mark=*]{(x * \tt) / ((\nn * x * sqrt((8000000000/(x*log2(\p)))) + (x + 1) * \tt * sqrt((8000000000/(x*log2(\p))))) * log2(\q))};
    \addlegendentry{SimplePIR (amortized)}

    \def\tt{1}
    
    \addplot [purple, thick, mark=*]{(x * \tt) / ((\nn * x * sqrt((8000000000/(x*log2(\p)))) + (x + 1) * \tt * sqrt((8000000000/(x*log2(\p))))) * log2(\q))};
    \addlegendentry{SimplePIR (not amortized)}
    
  \end{axis}
\end{tikzpicture}}
    \caption{Size of files (in bits) vs. rate for a database of size $1$GB}
  \end{subfigure}
  \caption{Comparison of rate for different database configurations}
\end{figure}
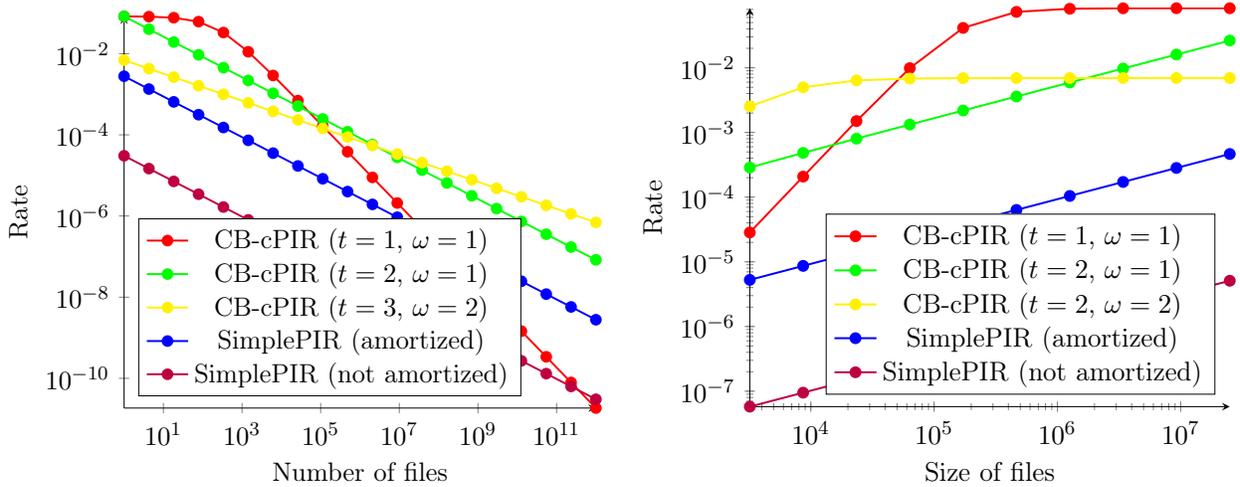

In the figures above, we compare the PIR rate of CB-cPIR with parameters $(q, n, k, s, v) = (2^{32}-5, 120, 60, 6, 4)$ against the PIR rate of SimplePIR, with parameters $(q, p, n) = (2^{32}, 495, 1024)$. Both protocols when instantiated with the above parameters offer $128$-bit security. The SimplePIR protocol considers a square database and is most appropriately compared with CB-cPIR over a square database (\emph{i.e.} $t=2, \omega = 1$). 

The SimplePIR communication costs are amortized over $100$ queries. 
In both scenarios, we see that CB-cPIR (on a square database) has favourable communication overhead in comparison with SimplePIR.

\subsubsection{Computational complexity} 

In this section we concretely determine the computational costs associated with the private retrieval of a file of size $L$ in $\mathbb{Z}_p$ using the SimplePIR protocol.

\textbf{Server complexity:} The SimplePIR protocol consists of a offline server preprocessing phase and an online query phase, the computational costs incurred by the server for these two phases are:
\begin{itemize}
    \item Preprocessing: This involves the cost of multiplying the matrices $DB \in \mathbb{Z}_p^{L\sqrt{m}\times \sqrt{m}}$ with $A \in \mathbb{Z}_q^{\sqrt{m} \times n}$, which amounts to $2Lmn$ operations over $\mathbb{Z}_q$.

    \item Per-query: This involves the cost of multiplying the database with a query $Q \in \mathbb{Z}_q^{\sqrt{m}\times L}$, which amounts to $2Lm$ operations in $\mathbb{Z}_q$.
\end{itemize}

\textbf{User complexity:} The user complexity for the SimplePIR protocol is divided into two parts, the complexity of generating the query and the complexity of decoding the server response.
\begin{itemize}
    \item Query generation: The complexity of query generation is dominated by the computation of $A\cdot s$, which requires $L\sqrt{m}n$ operations over $\mathbb{Z}_q$.
    \item Decoding response: The decoding of the response is dominated by the computation of $\mathrm{hint} \cdot s$, which requires $L\sqrt{m}n$ operations over $\mathbb{Z}_q$.
\end{itemize}

\begin{figure}[H]
  \centering
  \begin{subfigure}{0.49\textwidth}
    \centering
    \resizebox{\textwidth}{!}{\begin{tikzpicture}
  \begin{axis}[
    xlabel={Number of files},
    ylabel={Computational complexity},
    axis lines = left,
    legend pos = north west,
    domain=1:10000000000,
    xmode=log,
    ymode=log,
    ymax = 10^25,
    samples=10
    ]

    \def\L{32000}    
    \def\t{1}    
    \def\omegaw{1}    
    \def\w{2}    
    \def\n{120}  
    \def\s{6}    
    \def\d{120}  
    \def\k{60}    
    \def\v{1}    
    \def\tt{100}    
    \def\nn{1024}  
    \def\p{495}  
    \def\q{4294967296}  
    
    \addplot[red, thick, mark=*]{
    2 * (\L/(\d*32)) * x * \n * \s * ( ((2 * \n * \s) / (\d * x^(1/\t)))^\omegaw - 1 ) / ( ((2 * \n * \s) / (\d * x^(1/\t))) - 1 )
    + 2 * \omegaw * ( x^(1/\t) * \d * \k * \n * \s + x^(1/\t) * \d * \k * \v * \s + \d * \k * \w * \s + x^(1/\t) * \d * \k * \s ) 
    + ( 2 * \omegaw * (\L/(\d*32)) / \d ) * x^((\t-1)/\t) * \k * ( ((\n * \s) / (\d * x^(1/\t)))^\omegaw - 1 ) / ( ((\n * \s) / (\d * x^(1/\t))) - 1 ) * ( \n * \s + \s^2 + \w + \d * \s )
    };
    \addlegendentry{CB-cPIR ($t = 1$, $\omega = 1$)}
    
    \def\t{2}    
    \def\omegaw{1}    
    \addplot[green, thick, mark=*]{
    2 * (\L/(\d*32)) * x * \n * \s * ( ((2 * \n * \s) / (\d * x^(1/\t)))^\omegaw - 1 ) / ( ((2 * \n * \s) / (\d * x^(1/\t))) - 1 )
    + 2 * \omegaw * ( x^(1/\t) * \d * \k * \n * \s + x^(1/\t) * \d * \k * \v * \s + \d * \k * \w * \s + x^(1/\t) * \d * \k * \s ) 
    + ( 2 * \omegaw * (\L/(\d*32)) / \d ) * x^((\t-1)/\t) * \k * ( ((\n * \s) / (\d * x^(1/\t)))^\omegaw - 1 ) / ( ((\n * \s) / (\d * x^(1/\t))) - 1 ) * ( \n * \s + \s^2 + \w + \d * \s )
    };
    \addlegendentry{CB-cPIR ($t = 2$, $\omega = 1$)}

    \addplot[blue, thick, mark=*]{
     (2 * \L/(log2(\p)) * x * \nn + 2 * (\L/(log2(\p)) * x)*\tt + (2 * (\L/(log2(\p)) * sqrt(x) * \nn)*\tt))/\tt
    };
    \addlegendentry{SimplePIR (amortized)}

    \def\tt{1} 
    
    \addplot[purple, thick, mark=*]{
     (2 * \L/(log2(\p)) * x * \nn + 2 * (\L/(log2(\p)) * x)*\tt + (2 * (\L/(log2(\p)) * sqrt(x) * \nn)*\tt))/\tt
    };
    \addlegendentry{SimplePIR (not amortized)}
    
  \end{axis}
\end{tikzpicture}}
    \caption{Number of files vs. computational complexity for database with files of size $1$KB}
  \end{subfigure}
  \hfill
  \begin{subfigure}{0.49\textwidth}
    \centering
    \resizebox{\textwidth}{!}{\begin{tikzpicture}
  \begin{axis}[
    xlabel={Number of files},
    ylabel={Computational complexity},
    axis lines = left,
    legend pos = north west,
    domain=1:2500000,
    xmode=log,
    ymode=log,
    ymin = 4000000000,
    ymax = 200000000000000,
    samples=20
    ]

    \def\t{1}    
    \def\omegaw{1}    
    \def\w{2}    
    \def\n{120}  
    \def\s{6}    
    \def\d{120}  
    \def\k{60}    
    \def\v{1}    
    \def\tt{100}    
    \def\nn{1024}  
    \def\L{8000}  
    \def\p{495}  
    \def\q{4294967296}  
    
    \addplot[red, thick, mark=*]{
    2 * (8000000000/(3200*x)) * x * \n * \s * ( ((2 * \n * \s) / (\d * x^(1/\t)))^\omegaw - 1 ) / ( ((2 * \n * \s) / (\d * x^(1/\t))) - 1 )
    + 2 * \omegaw * ( x^(1/\t) * \d * \k * \n * \s + x^(1/\t) * \d * \k * \v * \s + \d * \k * \w * \s + x^(1/\t) * \d * \k * \s ) 
    + ( 2 * \omegaw * (8000000000/(3200*x))) * x^((\t-1)/\t) * \k * ( ((\n * \s) / (\d * x^(1/\t)))^\omegaw - 1 ) / ( ((\n * \s) / (\d * x^(1/\t))) - 1 ) * ( \n * \s + \s^2 + \w + \d * \s )
    };
    \addlegendentry{CB-cPIR ($t = 1$, $\omega = 1$)}
    
    \def\t{2}    
    \def\omegaw{1}    
    \addplot[green, thick, mark=*]{
    2 * (8000000000/(3200*x)) * x * \n * \s * ( ((2 * \n * \s) / (\d * x^(1/\t)))^\omegaw - 1 ) / ( ((2 * \n * \s) / (\d * x^(1/\t))) - 1 )
    + 2 * \omegaw * ( x^(1/\t) * \d * \k * \n * \s + x^(1/\t) * \d * \k * \v * \s + \d * \k * \w * \s + x^(1/\t) * \d * \k * \s ) 
    + ( 2 * \omegaw * (8000000000/(3200*x))) * x^((\t-1)/\t) * \k * ( ((\n * \s) / (\d * x^(1/\t)))^\omegaw - 1 ) / ( ((\n * \s) / (\d * x^(1/\t))) - 1 ) * ( \n * \s + \s^2 + \w + \d * \s )
    };
    \addlegendentry{CB-cPIR ($t = 2$, $\omega = 1$)}
    
    \addplot[blue, thick, mark=*]{
     ((2 * (8000000000/(log2(\p)*x)) * x * \nn) + (2 * (8000000000/(log2(\p)*x)) * x)*\tt + (2 * (8000000000/(log2(\p)*x)) * sqrt(x) * \nn)*\tt)/\tt
    };
    \addlegendentry{SimplePIR (amortized)}

    \def\tt{1}
    
    \addplot[purple, thick, mark=*]{
     ((2 * (8000000000/(log2(\p)*x)) * x * \nn) + (2 * (8000000000/(log2(\p)*x)) * x)*\tt + (2 * (8000000000/(log2(\p)*x)) * sqrt(x) * \nn)*\tt)/\tt
    };
    \addlegendentry{SimplePIR (not amortized)}
  \end{axis}
\end{tikzpicture}}
    \caption{Number of files vs. computational complexity for a database of size $1$GB}
  \end{subfigure}
  \caption{Comparison of computational complexity for different database configurations}
\end{figure}
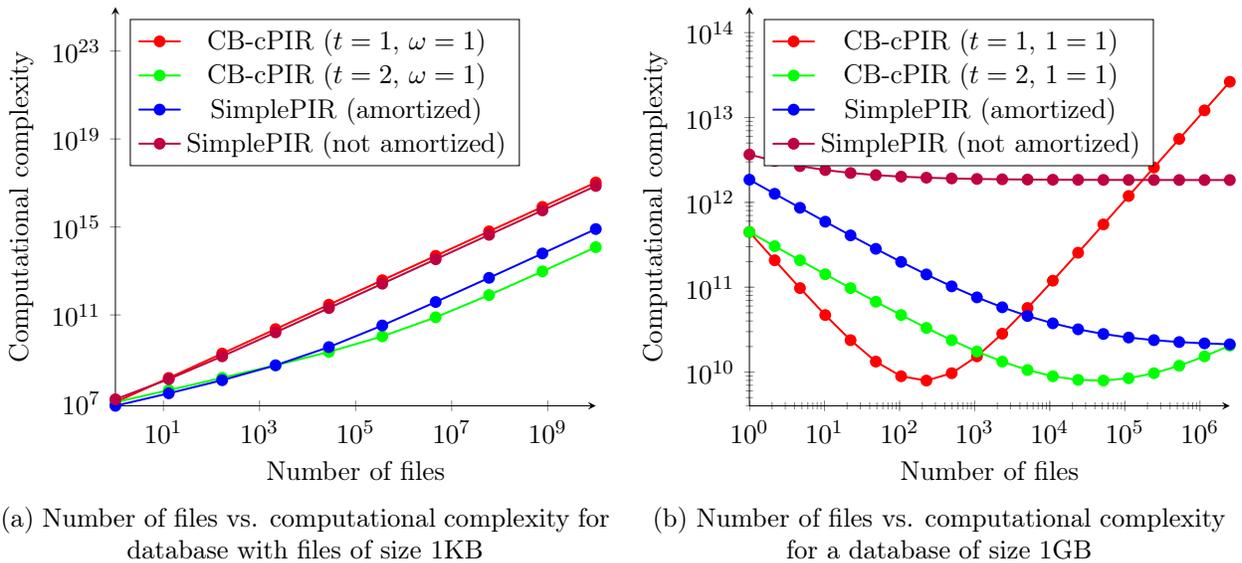

In the figures above, we compare the PIR rate of CB-cPIR with parameters $(q, n, k, s, v) = (2^{32}-5, 120, 60, 6, 4)$ against the PIR rate of SimplePIR, with parameters $(q, p, n) = (2^{32}, 495, 1024)$. Importantly, the modulus $q$ in both protocols is comparable in size ($\approx 32$ bits), which makes our comparisons valid. The SimplePIR protocol considers a square database and is most appropriately compared with CB-cPIR over a square database (\emph{i.e.} $t=2, \omega = 1$).

In both scenarios, we see that CB-cPIR (on a square database) has lower computational complexity in comparison with SimplePIR.


\section{Conclusions}
\label{sec:conclusion}
In this work, we present CB-cPIR, a code-based alternative for computational private information retrieval. Through a comprehensive comparison with state-of-the-art lattice-based schemes, we show that CB-cPIR is concretely cheaper in both communication and computational costs. These concrete advantages, combined with the scheme's structural simplicity, make CB-cPIR a practical and scalable solution for real-world PIR applications. Our results highlight the potential of code-based cryptography as a compelling direction for efficient computational private information retrieval.

Future work involves building a proof-of-concept implementation of CB-cPIR and exploring techniques to further reduce costs --- for example, by preprocessing the database.

\section*{Acknowledgments}

The authors would like to thank \c{S}. Bodur, R. Freij-Hollanti, E. Mart\'inez-Moro, and D. Ruano for useful discussions.


\bibliographystyle{plain}
\bibliography{bibfiles/cPIR,bibfiles/IT_PIR}

\end{document}